\documentclass[10pt,aps,prd,floatfix,notitlepage,showpacs,nofootinbib,superscriptaddress]{revtex4-1}

\usepackage{latexsym}
\usepackage{epsfig}
\usepackage{graphicx}
\usepackage{epstopdf}
\usepackage{amssymb}
\usepackage{amsmath}
\usepackage{dcolumn}
\usepackage{bm}
\usepackage{color}
\usepackage{comment}

\graphicspath{{pictures/}}

\begin{document}

\title{Interacting quintessence from a variational approach\\ Part I: algebraic couplings}

\author{Christian G. B\"ohmer}
\email{c.boehmer@ucl.ac.uk}
\affiliation{Department of Mathematics, University College London, Gower Street, London, WC1E 6BT, UK}

\author{Nicola Tamanini}
\email{nicola.tamanini@cea.fr}
\affiliation{Institut de Physique Th{\'e}orique, CEA-Saclay, F-91191, Gif-sur-Yvette, France}

\author{Matthew Wright}
\email{matthew.wright.13@ucl.ac.uk}
\affiliation{Department of Mathematics, University College London, Gower Street, London, WC1E 6BT, UK}


\begin{abstract}
We present a new approach to build models of quintessence interacting with dark or baryonic matter. We use a variational approach for relativistic fluids to realize an effective description of matter fields at the Lagrangian level. The coupling is introduced directly in the action by considering a single function mixing the dynamical degrees of freedom of the theory. The resulting gravitational field equations are derived by variations with respect to the independent variables. New interesting phenomenology can be obtained at both small scales, where new screening mechanisms for scalar fields can be realized, and large scales, where one finds an original and rich class of interacting quintessence models. The background cosmology of two of these models is studied in detail using dynamical system techniques. We find a variety of interesting results: for instance, these models contain dark energy dominated late time attractors and scaling solutions, both with early time matter dominated epochs and a possible inflationary origin. In general this new approach provides the starting point for future in depth studies on new interacting quintessence models.
\end{abstract}

\maketitle

\tableofcontents

\section{Introduction}

The universe is currently experiencing an epoch of accelerated expansion. During the last fifteen years, since it was first suggested by type-Ia supernovae surveys \cite{Riess:1998cb,Perlmutter:1998np}, this important conclusion has been confirmed by cosmological observations of ever increasing precision, which includes measurements of the cosmic microwave background (CMB) \cite{Komatsu:2010fb,Ade:2013zuv}, the Hubble constant \cite{Riess:2009pu}, baryon acoustic oscillations \cite{Lampeitl:2009jq} and again type-Ia supernovae \cite{Kowalski:2008ez}. Unfortunately, despite all these inputs from astronomical observations, on the theoretical ground we still lack a fully satisfactory explanation of this phenomenon. The standard cosmological model, the simplest model capable of fitting all the present observational data, assumes the existence of both a non-vanishing cosmological constant and a cold dark matter component. The former is introduced to produce the required accelerated expansion at late times, while the latter is postulated in order to increase the amount of structure formation needed to be in agreement with observed cosmological structures at both large and small scales. The cosmological constant, however, suffers from some profound theoretical issues arising from its extremely small measured value if compared with predicted values by (quantum) field theoretical considerations; see \cite{Martin:2012bt} for a recent review.

In order to solve these issues, or at least to alleviate them, it has been proposed that the current cosmological acceleration might be due to the effect of some dynamical field, mimicking the properties of a cosmological constant at late times. To agree with current observations such new component must be invisible to electro-magnetic radiation, which is the reason why it has been called dark energy, in analogy with dark matter. The fact that both dark matter and dark energy cannot be detected through the visible sector could imply new interesting phenomenology. In fact, as long as we are unable to probe these dark components but through their gravitational effects, one cannot exclude the possibility of non-gravitational interactions between them. Following this reasoning, one is thus naturally led to consider cosmological theories where interactions are possible. Note however that if both dark matter and dark energy can be measured solely by their gravitational influence, then the physical consequences of their interaction might be observationally indistinguishable from similar effects arising in other cosmological models without interaction, as for example in modified gravity and warm dark matter models \cite{Kunz:2007rk,Wei:2008vw,Wei:2013rea}. Nevertheless it is important to study different interacting dark energy models not only to search for new interesting phenomenological features, but also to support the observational efforts to focus on possible distinctive signatures of such models; a tentative review on these models has recently been proposed in \cite{Bolotin:2013jpa} (see also \cite{Timothy:2013iia}).

The simplest dynamical models of dark energy identify this mysterious component with a minimally coupled canonical scalar field $\phi$. Dark energy theories of this kind are generally known by the name of quintessence and are quite popular in the literature because they are sufficiently simple to handle and sufficiently complicated to produce non-trivial dynamics; for reviews see e.g.~\cite{Copeland:2006wr,Tsujikawa:2013fta}.

On the other hand, the cosmological evolution of (cold) dark matter is usually described by a perfect fluid with energy density $\rho$ and vanishing pressure $p=0$. Simple models of interacting dark energy can thus be constructed by coupling the scalar field $\phi$ to the energy density $\rho$ of the dark matter fluid. Such a coupling is usually introduced at the level of the cosmological field equations adding a term mixing the quintessence and dark matter equations of motion. Specifically in the presence of a general coupling between the two dark components, the background cosmological equations are commonly expressed as
\begin{align}
  3H^2 = \kappa^2 \left(\rho + \rho_\phi\right) \,, \qquad \dot{\rho} + 3H \rho = -Q \qquad \dot{\rho_\phi} + 3H \left(\rho_\phi+p_\phi\right) = Q \,,
  \label{eq:standard_coupling}
\end{align}
where $H$ is the Hubble rate, $\rho_\phi$ the energy density of quintessence, $p_\phi$ its pressure and $\kappa^2=8\pi G$. We set $c=1$. An over-dot denotes differentiation with respect to cosmological time and the variable $Q$ quantifies the rate of energy exchanged in the dark sector and generally depends on both $\phi$ and $\rho$ (and possibly their time derivatives).
If $Q>0$ the energy flows from dark matter to quintessence, while if $Q<0$ the energy transfer is in the opposite direction. 
The dependence on $\phi$ and $\rho$ of $Q$ is a phenomenological assumption which, together with the form of the scalar field potential, determines the interacting quintessence model at hand.
Throughout the last fifteen years an impressive amount of possible choices has been studied, the most common being $Q\propto\rho\dot\phi$, which can easily be motivated in the Einstein frame formulation of scalar-tensor theories \cite{Amendola:1999er,Amendola:1999qq,Holden:1999hm,Wetterich:1994bg}. We cannot discuss in detail this vast amount of literature and the interested reader is referred to \cite{Bolotin:2013jpa,Timothy:2013iia} and in the references therein which provides a general overview of this subject (see also \cite{Shojai:2013gsa,Tzanni:2014eja,Fadragas:2014mra,Costa:2014pba} for some recent works).

In the present paper we will consider a new approach to build models of quintessence interacting with dark matter. The coupling will not be added at the level of the cosmological field equations as in Eqs.~(\ref{eq:standard_coupling}), but it will be directly introduced into an action characterising both quintessence and dark matter. Variations with respect to the scalar field $\phi$ and the matter degrees of freedom will then provide the coupled equations of motion of dark matter and dark energy. We will assume that dark matter can still effectively be characterised by a perfect fluid with vanishing pressure even at the Lagrangian level. This implies that a Lagrangian description for perfect fluids must be adopted in order not only to obtain the correct expression for the dark matter energy-momentum tensor, but also to define its coupling to the quintessence field.

There are different Lagrangian approaches to perfect fluids that have been developed during the years; see e.g.~\cite{Taub:1954zz,Schutz:1970my,Schutz:1977df,Bailyn:1980zz,Brown:1992kc,Andersson:2006nr}. Pourtsidou {\it et al} \cite{Pourtsidou:2013nha} have recently considered a Lagrangian approach to a scalar field coupled to a fluid based on the so-called pull-back formalism~\cite{Andersson:2006nr}. In this framework they constructed three new models of interacting dark energy with completely new features with respect to the models defined by the standard coupling of Eqs.~(\ref{eq:standard_coupling}). Subsequently they analysed their impact on the cosmological observables showing new interesting phenomenology capable of being tested against observational data. The same approach has also been used to study theories of interacting cosmological fluids in the context of effective field theories \cite{Ballesteros:2013nwa}.

In this paper we will develop the Lagrangian formulation of quintessence coupled to the dark matter fluid. Instead of using the pull-back formalism, we will employ the framework with Lagrangian multipliers outlined by Brown in \cite{Brown:1992kc}, which is easier to handle and will lead to new phenomenology at both Solar System and cosmological scales. Our philosophy will be to fully embrace the variational methods of \cite{Brown:1992kc} and in particular we will discuss the most general actions that one can build out of all the dynamical degrees of freedom associated with the matter fluid. A similar approach has already been studied for dark matter non-minimally coupled to gravity where the fluid 4-velocity is coupled to the Ricci tensor and the fluid energy density to the Ricci scalar \cite{Bettoni:2011fs,Bettoni:2015wla}.
Here however we will not modify the gravitational sector, which will be described by standard general relativity, but we will limit the interaction of the matter fluid with the quintessence field. The couplings obtained in this way will first be applied to describe the phenomenology of dark energy at small scales, where models presenting environment dependent interactions will be able to efficiently screen any scalar field effect and thus to satisfy all Solar System constraints. Then the background cosmology obtained within this framework will be studied and the dynamics of different interacting quintessence models will be analysed in depth.

The paper is organised as follows. In Sec.~\ref{sec:fluid_Lagrangian} we will briefly review the variational set up for relativistic fluids exposed in \cite{Brown:1992kc}. In Sec.~\ref{sec:algebraic_coupling} this variational approach will be employed to build theories of a perfect fluid interacting with a scalar field. The most general algebraic coupling will be considered, deriving the equations of motion and discussing the conservation equations. In Sec.~\ref{sec:small_scales} this new way of coupling a relativistic fluid with a scalar field will be used to build screening mechanisms for dark energy capable of passing every experimental test and Solar System observation. Sec.~\ref{sec:cosmology} will then be dedicated to cosmology: the cosmological equations will be obtained and the evolution of the universe will be analysed with dynamical system techniques. Finally we will draw conclusions and discuss future perspectives in Sec.~\ref{sec:conclusion}.

{\bf Notation and conventions}: Unless otherwise specified we will assume standard general relativistic notation with signature $(-,+,+,+)$ and Greek indices running from 0 to 3. Sometimes the comma notation for partial derivatives will be used: for example $\phi_{,\mu}=\partial_\mu\phi$. Units where $c=\hbar=1$ will be employed together with $\kappa^2=M_{\rm P}^{-2}=8\pi G$.

\section{Lagrangian formulation of relativistic fluids}
\label{sec:fluid_Lagrangian}

In this section we outline the variational approach to relativistic fluids following closely the formulation of Brown. We will only introduce the Lagrangian and derive the equations of motion while skipping the detailed discussions regarding the thermodynamical features that can be obtained in this context. The reader interested in more information is referred to \cite{Brown:1992kc}.

Within Brown's framework the Lagrangian for the relativistic fluid can be written as
\begin{align}
\mathcal{L}_M = -\sqrt{-g}\,\rho(n,s) + J^\mu\left(\varphi_{,\mu}+s\theta_{,\mu}+\beta_A\alpha^A_{,\mu}\right) \,,
\label{001}
\end{align}
where $g$ is the determinant of the metric tensor $g_{\mu\nu}$ and $\rho$ is the energy density of the fluid. We assume $\rho(n,s)$ to be prescribed as a function of $n$, the particle number density, and $s$, the entropy density per particle.
$\varphi$, $\theta$ and $\beta_A$ are all Lagrange multipliers with $A$ taking the values $1,2,3$ and $\alpha_A$ are the Lagrangian coordinates of the fluid. The vector-density particle number flux $J^\mu$ is related to $n$ as
\begin{align}
J^\mu=\sqrt{-g}\,n\,U^\mu\,, \qquad |J|=\sqrt{-g_{\mu\nu}J^\mu J^\nu}\,, \qquad n=\frac{|J|}{\sqrt{-g}} \,,
\label{056}
\end{align}
where $U^\mu$ is the fluid 4-velocity satisfiyng $U_\mu U^\mu=-1$. 

The independent dynamical variables which have to be considered in the variation of the Lagrangian (\ref{001}) are $g^{\mu\nu}$, $J^\mu$, $s$, $\varphi$, $\theta$, $\beta_A$ and $\alpha^A$. 

Variation with respect to the metric $g^{\mu\nu}$ gives the following energy-momuntum tensor
\begin{align}
T_{\mu\nu} = \rho\,U_\mu U_\nu +\left(n\frac{\partial\rho}{\partial n}-\rho\right) \left(g_{\mu\nu}+U_\mu U_\nu\right) \,,
\end{align}
which can be rewritten in the standard perfect fluid form
\begin{align}
T_{\mu\nu} = p\, g_{\mu\nu} + (\rho+p)\, U_\mu U_\nu \,,
\label{015}
\end{align}
once the pressure $p$ has been identified with
\begin{align}
p = n\frac{\partial\rho}{\partial n}-\rho \,.
\end{align}
Variation with respect to the other variables gives
\begin{align}
J^\mu:& \qquad \mu\,U^\mu + \varphi_{,\mu}+s\theta_{,\mu}+\beta_A\alpha^A_{,\mu} =0\,,\label{002}\\
s:& \qquad -\frac{\partial\rho}{\partial s}+n\,U^\mu\theta_{,\mu} =0 \,,\label{003}\\
\varphi:& \qquad J^\mu{}_{,\mu}=0 \,,\label{004}\\
\theta:& \qquad (sJ^\mu)_{,\mu}=0 \,,\label{005}\\
\beta_A:& \qquad J^\mu\alpha^A_{,\mu}=0 \,,\label{006}\\
\alpha^A:& \qquad (J^\mu\beta_A)_{,\mu}=0 \label{007}\,,
\end{align}
where 
\begin{align}
\mu = \frac{\rho+p}{n} = \frac{\partial\rho}{\partial n} \,,
\label{011}
\end{align}
is the chemical potential.
Eqs.~(\ref{004}) and (\ref{005}) stand for the particle number conservation constraint and the entropy exchange constraint, respectively. These can be rewritten also as
\begin{align}
\nabla_\mu(n\,U^\mu)=0 \quad{\rm and}\quad \nabla_\mu(s\,n\,U^\mu)=0 \,,
\label{019}
\end{align}
where $\nabla_\mu$ is the covariant derivative with respect to $g_{\mu\nu}$. The scalar fields $\varphi$ and $\theta$ are thus Lagrange multipliers which impose these two constraints. In the same manner $\beta_A$ are three Lagrange multipliers which are needed to restrict the fluid's 4-velocity vector to be directed along the flow lines with constant $\alpha^A$, as specified by Eq.~(\ref{006}). The meaning of Eq.~(\ref{003}) can be understood by recalling from the first principle of thermodynamics. If the particle number is conserved one can identify the temperature $T$ with
\begin{align}
T = \frac{1}{n}\frac{\partial\rho}{\partial n} \,.
\end{align}
From Eq.~(\ref{003}) we thus have that $\theta$ behaves as a potential for the temperature
\begin{align}
T = U^\mu\theta_{,\mu} \,.
\label{008}
\end{align}
Finally Eq.~(\ref{007}) determines the dynamics of the Lagrange multipliers $\beta_A$, while Eq.~(\ref{002}) is known as the potential representation of the fluid's 4-velocity. Combining this equation with Eqs.~(\ref{003}) and (\ref{006}) we find
\begin{align}
F = \mu - T\,s = U^\mu\varphi_{,\mu} \,,
\label{014}
\end{align}
where $F$ is the chemical free energy. We thus obtain an analogy with Eq.~(\ref{008}) implying that $\varphi$ is a potential for the chemical free energy.

It is possible to show that the fluid field equations
\begin{align}
\nabla_\mu T^{\mu\nu} =0 \,,
\label{009}
\end{align}
are indeed implied by Eqs.~(\ref{002})-(\ref{007}). To prove this one can follow the calculations outlined in Appendix~\ref{appA} assuming no interaction with the scalar field; see also \cite{Brown:1992kc}. Here we just mention that contracting Eq.~(\ref{009}) with the projective tensor
\begin{align}
h_{\mu\nu} = g_{\mu\nu} +U_\mu U_\nu \,,
\label{017}
\end{align}
gives the equations
\begin{align}
\frac{d U^\mu}{d\lambda} +\Gamma^\mu_{\alpha\beta} U^\alpha U^\beta = \frac{p_{,\nu}}{\rho+p}h^{\nu\mu} \,,
\label{eq:geodesic_standard}
\end{align}
which reduces to the usual geodesic equation for a pressure-less fluid (dust); see for example \cite{Wald:1984rg}.

To conclude this section we mention that once equations (\ref{002})-(\ref{007}) are imposed, one can obtain the on-shell Lagrangian by substituting them back into (\ref{001}). This procedure yields
\begin{align}
\mathcal{L}_M = -\sqrt{-g} \, \rho =\sqrt{-g}\, p \quad \mbox{(on-shell)}\,,
\end{align}
with the last equivalence holding up to total derivatives.
This means that the on-shell Lagrangian can be represented by either negative energy density, or simply by the pressure.
This result does not hold however if the fluid is coupled to other fields or non-minimally coupled to gravity \cite{Faraoni:2009rk,Harko:2010zi,Minazzoli:2012md}.

\section{Relativistic fluid interacting with a scalar field: algebraic coupling}
\label{sec:algebraic_coupling}

The standard way of coupling two interacting matter components in general relativity consists in adding a non-vanishing current $Q_\mu$ to the right hand side of the conservation equations.
If $T^{(A)}_{\mu\nu}$ and $T^{(B)}_{\mu\nu}$ are the energy-momentum of these two matter components, then their conservation equations can be written as
\begin{align}
	\nabla^\mu T^{(A)}_{\mu\nu} = Q_\nu \qquad\mbox{and}\qquad \nabla^\mu T^{(B)}_{\mu\nu} = -Q_\nu \,,
	\label{032}
\end{align}
while the Einstein field equations are still given by
\begin{align}
	G_{\mu\nu} = \kappa^2 \left( T^{(A)}_{\mu\nu} + T^{(B)}_{\mu\nu} \right) \,.
	\label{033}
\end{align}
This applies also to the case in which one of the two components is the dark matter fluid while the other is a scalar field, namely quintessence.

In what follows, instead of introducing the interaction at the level of the field equations as above, we will employ the variational methods developed in Sec.~\ref{sec:fluid_Lagrangian} in order to define a coupling between a scalar field and a perfect fluid at the Lagrangian level.

\subsection{Lagrangian formulation and field equations}
\label{sec:algebraic_formulation}

Taking into account the formulation of Sec.~\ref{sec:fluid_Lagrangian}, we can now set up our model where a scalar field $\phi$ interacts with a matter fluid. The action we will consider is
\begin{align}
\mathcal{S} = \int d^4x \left(\mathcal{L}_{\rm grav} +\mathcal{L}_M+ \mathcal{L}_\phi+ \mathcal{L}_{\rm int}\right) \,,
\label{010}
\end{align}
where the matter Lagrangian $\mathcal{L}_M$ represents a perfect fluid and is given in (\ref{001}); the gravitational sector $\mathcal{L}_{\rm grav}$ is given by the standard Einstein-Hilbert Lagrangian
\begin{align}
\mathcal{L}_{\rm grav} = \frac{\sqrt{-g}}{2\kappa^2}R \,,
\end{align}
where $R$ is the curvature scalar with respect to the metric $g_{\mu\nu}$; the scalar field Lagrangian is given by
\begin{align}
\mathcal{L}_\phi = -\sqrt{-g}\, \left[\frac{1}{2}\partial_\mu\phi\,\partial^\mu\phi +V(\phi)\right] \,,
\end{align}
with $V$ a general potential for $\phi$;
finally for the interacting sector we will consider a general algebraic coupling of the type
\begin{align}
\mathcal{L}_{\rm int} = -\sqrt{-g}\, f(n,s,\phi) \,, \label{027}
\end{align}
where $f(n,s,\phi)$ is an arbitrary function which will specify the particular model at hand.
Note that $f$ cannot depend on the Lagrange multipliers, which are needed to impose the constraints (\ref{003})--(\ref{007}) for the matter fluid and are not supposed to mix with the scalar field.
In other words, we are assuming that only the {\it dynamical} degrees of freedom of the fluid, namely $n$ and $s$, couple with the scalar field.
Note that in the Lagrangian approach of Sec.~\ref{sec:fluid_Lagrangian} the particle number $n$ is not treated as a fundamental variable to use in the variation, but only as a function composed by $J^\mu$ and $g_{\mu\nu}$ according to Eqs.~(\ref{056}).
One can thus also consider a more general coupling between the variables $J^\mu$, $s$ and $\phi$.
However if no derivatives are assumed to enter the interacting term, then the function (\ref{027}) still represents the most general coupling one can build out of the variables $J^\mu$, $s$ and $\phi$.
In the first part of the present work (namely the present paper) we will only consider {\it algebraic} couplings of the kind (\ref{027}) where no derivatives appear.
Interacting terms with derivatives will be discussed in the second part \cite{part2}.

Action~(\ref{010}) has to be varied with respect to the fields $g^{\mu\nu}$, $\phi$, $J^\mu$, $s$, $\varphi$, $\theta$, $\beta_A$ and $\alpha^A$. However the variation with respect to the last four among these will give again Eqs.~(\ref{004})-(\ref{007}) and thus will not be repeated.

We start considering the variation in $s$ which produces the following equation
\begin{align}
U^\mu\theta_{,\mu} = \frac{1}{n}\frac{\partial\rho}{\partial n}+\frac{1}{n}\frac{\partial f}{\partial n} = T + T_{\rm int} \,.
\label{012}
\end{align}
The scalar field $\theta$ represents now a potential not only for the fluid temperature $T$ but also for the ``interacting temperature'' $T_{\rm int}$. Variation with respect to $J^\mu$ yields the equation
\begin{align}
(\mu+\mu_{\rm int})\,U_\mu+ \varphi_{,\mu}+s\theta_{,\mu}+\beta_A\alpha^A_{,\mu} =0 \,,
\label{013}
\end{align}
where $\mu_{\rm int}=\partial f/\partial n$ has been defined in analogy with the chemical potential (\ref{011}). This equation modifies the potential representation of the fluid 4-velocity (\ref{002}). Using Eqs.~(\ref{006}), (\ref{012}) and (\ref{013}) we can show that
\begin{align}
U^\mu\varphi_{,\mu} = F + F_{\rm int} \,,
\label{031}
\end{align}
where $F_{\rm int}=\mu_{\rm int}-s\,T_{\rm int}$. This equation generalizes Eq.~(\ref{014}) and tells us that $\varphi$ is now a potential for the chemical free energy of the fluid plus its interacting counterpart.
In general Eqs.~(\ref{012})--(\ref{031}), together with Eqs.~(\ref{004})-(\ref{007}), define how the thermodynamical properties of the fluid change due to the interaction with the scalar field. We will not discuss these relations further but focus on the spacetime dynamics only.

At this point we perform the variation with respect to the metric tensor which results in the following Einstein field equations
\begin{align}
G_{\mu\nu} = \kappa^2 \left(T_{\mu\nu}+T_{\mu\nu}^{(\phi)} +T_{\mu\nu}^{\rm (int)}\right) \,,
\label{016}
\end{align}
where $T_{\mu\nu}$ is the fluid energy-momentum tensor given in (\ref{015}) and
\begin{align}
T_{\mu\nu}^{(\phi)} = \partial_\mu\phi\,\partial_\nu\phi -g_{\mu\nu} \left[\frac{1}{2}\partial_\mu\phi\,\partial^\mu\phi +V(\phi)\right] \,, \label{034}
\end{align}
is the energy-momentum tensor of the scalar field $\phi$. The interacting energy-momentum tensor $T_{\mu\nu}^{\rm (int)}$ can be written as
\begin{align}
T_{\mu\nu}^{\rm (int)} = p_{\rm int}\,g_{\mu\nu} + \left(p_{\rm int}+\rho_{\rm int}\right) U_\mu U_\nu \,,
\end{align}
where we have defined 
\begin{align}
\rho_{\rm int} = f(n,s,\phi) \quad\mbox{and}\quad p_{\rm int} = n\frac{\partial f(n,s,\phi)}{\partial n}-f(n,s,\phi) \,,
\label{024}
\end{align}
as the interacting energy density and pressure. Finally the variation with respect to the scalar field yields the following modified Klein-Gordon equation
\begin{align}
\Box\phi-\frac{\partial V}{\partial\phi}-\frac{\partial f}{\partial\phi} =0 \,
\label{018}
\end{align}
where $\Box= \nabla^\mu \nabla_\mu$.

\subsection{Conservation equations}

It is interesting to reformulate the field equations above as in the standard approach to coupled matter component in general relativity, i.e.~as defined by Eqs.~(\ref{032})--(\ref{033}).
For this purpose one can define a new energy-momentum tensor for the fluid as $\tilde{T}_{\mu\nu} = T_{\mu\nu} + T_{\mu\nu}^{\rm (int)}$ with the energy density $\tilde\rho = \rho + \rho_{\rm int}$ and pressure $\tilde{p} = p + p_{\rm int}$.
In this case the Einstein field equations (\ref{016}) become
\begin{align}
	G_{\mu\nu} = \kappa^2 \left(\tilde{T}_{\mu\nu}+T_{\mu\nu}^{(\phi)}\right)\,,
\end{align}
resembling Eqs.~(\ref{033}).
Using the Klein-Gordon equation (\ref{018}) and the energy-momentum tensor (\ref{034}), the conservation equation for the scalar field can be written as
\begin{align}
	\nabla^\mu T_{\mu\nu}^{(\phi)} = \frac{\partial\tilde\rho}{\partial\phi} \partial_\nu\phi = Q_\nu\,,
	\label{eq:scalar_conserv_cov_tilde}
\end{align}
which shows that the scalar field is not conserved due to the interaction with the fluid.

To prove that also the fluid energy-momentum is not conserved, it is better to split its conservation equation into the parallel and perpendicular components to the fluid flow:
\begin{align}
	\nabla^\mu \tilde{T}_{\mu\nu} = h^\lambda_\nu \nabla^\mu \tilde{T}_{\mu\lambda} - U_\nu U^\lambda \nabla^\mu \tilde{T}_{\mu\lambda} \,, \label{035}
\end{align}
where $h_{\mu\nu}$ is defined in Eq.~(\ref{017}).
Then using Eqs.~(\ref{004})-(\ref{007}) and Eq.~(\ref{013}) one can show that (see Appendix~\ref{appA})
\begin{align}
  h_\mu^\nu \nabla^\lambda \tilde{T}_{\lambda\nu} = - h_\mu^\nu \frac{\partial\tilde\rho}{\partial\phi} \nabla_\nu\phi\,,
  \qquad\mbox{and}\qquad
  U_\nu \nabla_\mu \tilde{T}^{\mu\nu} = - U^\nu \frac{\partial\tilde\rho}{\partial\phi} \nabla_\nu\phi \,.
  \label{051}
\end{align}
Inserting this back into Eq.~(\ref{035}) gives
\begin{align}
	\nabla^\mu \tilde{T}_{\mu\nu} = -\frac{\partial\tilde\rho}{\partial\phi} \partial^\nu\phi = -Q_\nu\,.
	\label{eq:matter_cov_tilde}
\end{align}
This shows that the Lagrangian formulation we are considering in the present paper can be mapped back into the standard relativistic approach defined by Eqs.~(\ref{032})--(\ref{033}).
Note however that the energy density and pressure of the fluid differ from the uncoupled case in as much as they now depend also on the scalar field through the interacting term.
In particular even if $p=0$, e.g.~in cold dark matter applications, the pressure $\tilde{p}$ is not expected to vanish.
As we will see these new features will give rise to a new interesting phenomenology at both small and large scales.

The argument above can also be repeated directly with $T_{\mu\nu}$ which does not depend on the scalar field.
This provides
\begin{align}
  U_\nu\nabla_\lambda T^{\lambda\nu} = 0  \qquad\mbox{and}\qquad h_{\mu\nu} \nabla_\lambda T^{\lambda\nu} = -2n U^\lambda \nabla_{[\lambda} \left( U_{\mu]} \mu_{\rm int} \right) \,,
  \label{036}
\end{align}
where square brackets between two indices denote antisymmetrisation.
Eq.~(\ref{036}) implies that in general
\begin{align}
  \nabla^\mu T_{\mu\nu} \neq 0 \,,
\end{align}
meaning that the uncoupled part of the matter fluid is no more conserved in the presence of the coupling to the scalar field. Note however that the longitudinal part of the conserved equation, the one aligned with $U^\mu$, is always conserved. This will be relevant for cosmology where the isotropy and homogeneity of the spacetime will imply the conservations of $T_{\mu\nu}$ at the background level.

\section{Small scale phenomenology}
\label{sec:small_scales}

Before employing the results developed so far to construct new models of quintessence coupled to dark matter, we study their possible applications to Solar System and galaxy scale phenomenology. For this purpose in this section we will use the Lagrangian approach developed in Sec.~\ref{sec:algebraic_formulation} to investigate new couplings between the scalar field $\phi$ and baryonic matter. In what follows thus $\rho$ and $p$ will represent the energy density and pressure of matter constituted by Standard Model particles.

\subsection{Fifth force}

We start recalling that because of the Bianchi identity the relation $\nabla_\mu G^{\mu\nu}=0$ holds.
We then expect the right hand side of Eq.~(\ref{016}) to be covariantly conserved, giving in this way the equations of motion of the matter fluid.
Taking the covariant derivative of Eq.~(\ref{016}), contracting with $h_{\mu\nu}$ as given in Eq.~(\ref{017}), and using the scalar field equation (\ref{018}), eventually yields
\begin{align}
\frac{d U^\mu}{d\lambda}+\Gamma^\mu_{\sigma\nu} U^\sigma U^\nu =-\frac{1}{\rho+p+p_{\rm int}+\rho_{\rm int}}\left(p_{,\nu}+p_{\rm int}{}_{,\nu}+\frac{\partial f}{\partial\phi}\phi_{,\nu}\right)h^{\nu\mu} \,.
\label{020bis}
\end{align}
which generalises Eq.~(\ref{eq:geodesic_standard}).
For a pressure-less fluid, or equivalently for point particles, this expression reduces to
\begin{align}
\frac{d U^\mu}{d\lambda}+\Gamma^\mu_{\sigma\nu} U^\sigma U^\nu =-\frac{1}{\rho+p_{\rm int}+\rho_{\rm int}}\left(p_{\rm int}{}_{,\nu}+\frac{\partial f}{\partial\phi}\phi_{,\nu}\right)h^{\nu\mu} = f^\mu \,.
\label{020}
\end{align}
Comparing Eqs.~(\ref{020}) and (\ref{eq:geodesic_standard}) one realises that the interaction between the scalar field and the matter fluid gives rise to a fifth force $f^\mu$.
Thus, as one could expect, the motion of the matter fluid is non-geodesic due to the presence of the extra force $f_\mu$ which in general depends on both $n$ and $\phi$.
Note however that if $f^\mu$ does not depend on either $\phi$ or $n$ then it vanishes completely, reflecting the fact that there is no effective interaction in such cases.
Furthermore, according to the definition of 4-force in relativistic theories, $f_\mu$ is always orthogonal to the fluid 4-velocity: $U^\mu f_\mu=0$.

The appearance of this fifth force can give rise to interesting phenomenology at both Solar System and galactic scales.
For example the non-geodesic motion of luminous matter can be used as an alternative to dark matter in explaining the anomalies observed in the galaxy rotation curves.
The situation here is similar to the one arising from non-minimally coupled theories \cite{Bertolami:2007gv}, where an extra force depending on the local curvature produces effects similar to the ones predicted by MOND theories \cite{Milgrom:1983ca}.
In MOND theories however the modification of Newtonian mechanics at galactic distances depends on an effective acceleration $a_M$ (or equivalently on an effective length $l_M$) which is usually assumed to be constant and thus difficult to adapt to different galaxy profiles.
In non-minimally coupled theories such effective acceleration depends on the small value of the local gravitational curvature as well as on the matter energy density of the environment.
This might help in fitting the rotational velocity curves of several galaxies within a single model where $a_M$ varies from one galaxy to the other.
Exactly the same argument applies to the scalar field interacting model considered here.
From the fifth force (\ref{020}) it is in fact possible to define an effective acceleration $a_\phi$ which modifies the Newtonian equations at galactic distances in exactly the same way as $a_M$ does.
The difference resides in the fact that $a_\phi$ depends on the scalar field $\phi$ rather than on the local curvature and thus might predict quite different results even for galaxies having roughly the same matter density distribution.

At Solar System scales the fifth force (\ref{020}) is highly constrained by the experiments which do not show any violation on the inverse-square Newtonian law up to parts in $10^9$ \cite{Will:2014xja}.
The effects of the interaction between the scalar field and the matter fluid must then be negligible at those distances.
Fortunately, as we are now going to see, the dependence of the fifth force on the fluid dynamical degrees of freedom allows us to build efficient screening of possible deviations from the geodesic motion.

\subsection{Chameleon mechanism}

From the Klein-Gordon equation (\ref{018}) one can realise that the scalar field $\phi$ ``feels'' an effective potential given by
\begin{align}
  V_{\rm eff} = V(\phi) + f(n,s,\phi) \,,
  \label{037}
\end{align}
which can be used to screen the interaction between the scalar field and matter at Solar System scales.
The effective potential (\ref{037}) represents in fact a generalisation of the so called chameleon theories where the scalar field couples to the matter energy density and acquire an effective mass characterised by the surrounding environment \cite{Khoury:2003aq}.
To show this we can consider an interacting term defined by
\begin{align}
f(n,s,\phi) = \rho(n,s)\, \gamma(\phi) \,,
\label{041}
\end{align}
where $\rho$ is the matter energy density and $\gamma$ is a function of $\phi$. With this choice the interacting energy density and pressure (\ref{024}) read
\begin{align}
\rho_{\rm int}=\rho(n,s)\, \gamma(\phi) \quad\mbox{and}\quad p_{\rm int}=p(n,s)\, \gamma(\phi) \,.
\label{021}
\end{align}
Next we assume the scalar field potential to be of the ``runaway kind'', i.e.~to satisfy
\begin{align}
\lim_{\phi\rightarrow\infty}V=0\,, \qquad \lim_{\phi\rightarrow\infty}\frac{\partial^{(i)}V}{\partial\phi^{(i)}}/ \frac{\partial^{(i-1)}V}{\partial\phi^{(i-1)}}=0 \quad\mbox{for }i=1,2,3,... \,,\label{039}\\
\lim_{\phi\rightarrow 0}V=\infty\,, \qquad \lim_{\phi\rightarrow 0}\frac{\partial^{(i)}V}{\partial\phi^{(i)}}/ \frac{\partial^{(i-1)}V}{\partial\phi^{(i-1)}}=\infty \quad\mbox{for }i=1,2,3,... \,.\label{040}
\end{align}
Such constraints are usually imposed on the scalar field potential in order to obtain a late time cosmic speed-up such as, for example, in the well-known inverse power law potential \cite{Zlatev:1998tr}
\begin{align}
V(\phi) = \frac{M^{4+\alpha}}{\phi^\alpha} \,,
\end{align}
where $M$ is a constant with mass units and $\alpha$ a positive parameter. If we now set
\begin{align}
\gamma(\phi) = \gamma_0 e^{\kappa\beta\phi} \,,
\end{align}
with $\beta$ and $\gamma_0$ positive constants, we obtain the effective potential
\begin{align}
V_{\rm eff} = V(\phi)+\gamma_0\,\rho\,e^{\kappa\beta\phi} \,,
\label{038}
\end{align}
which coincides with the single matter chameleon potential of \cite{Khoury:2003aq}. Of course if we considered more than one matter fluid we would have ended up with exactly the multi-component equation considered in \cite{Khoury:2003aq}, however for our scopes a discussion with one matter fluid is enough.
The fact that the $V_{\rm eff}$ has a minimum for Solar System values of $\rho$ implies that at those distances the scalar field acquires an effective mass.
If this mass is sufficiently high, then all possible interactions with particles of the Standard Model are efficiently suppressed.
At cosmological scales though, where the matter energy density is extremely low, the effective potential (\ref{038}) reduces simply to $V(\phi)$ which, given the assumptions (\ref{039}) and (\ref{040}), can thus drive the late time acceleration of the universe; see \cite{Khoury:2003aq} for more details.

From these considerations we have seen that the formulation developed in Sec.~\ref{sec:algebraic_formulation}, with the assumption $f(n,s,\phi) = \rho(n,s)\, \gamma(\phi)$, can be used to derive chameleon field theories from a Lagrangian approach.
Moreover if other interacting terms $f(n,s,\phi)$ are considered, the same formulation can also be adopted to build more general screening mechanisms for the scalar field.
In fact as long as the effective potential (\ref{037}) posses a minimum for Solar System values of $n$ and $s$ then the scalar field will always appear with a non-zero mass at those scales, which, if large enough, will allow the scalar field to pass all Solar System tests.
Of course, in order to properly check its viability, any such model should undergo an analysis similar to the one performed for the original chameleon theories of \cite{Khoury:2003aq}.
A deeper analysis of the possible screening mechanisms that can be obtained with different choices of the interacting term $f(n,s,\phi)$ will be left for future studies.
Here, in order to present some of the potentialities that derive from our approach, we will limit the discussion on few considerations regarding a particular model.

First we will assume the function $f(n,s,\phi)$ to depend on $n$ and $s$ only through $\rho(n,s)$.
This reduces the space of possible new models, but allows for an immediate physical interpretation and the analysis can be better compared with the standard chameleon paradigm as given by Eq.~(\ref{041}).
In any case the reader should keep in mind that from the Lagrangian formulation developed in Sec.~\ref{sec:algebraic_formulation} it is possible to construct screening mechanism where the mass of the scalar field depends on the local number density $n$ and entropy $s$.
Note that though the effects of $n$ might be similar to the one obtained with $\rho$, the possibility of creating ``entropic screenings'' using different values of $s$ is an interesting feature arising from our approach which might be worthy investigate in future works\footnote{One might also think to relax the condition on the conservation on the entropy density (\ref{005}) to probe possible non-adiabatic effects.}.

\begin{figure}[!ht]
\includegraphics{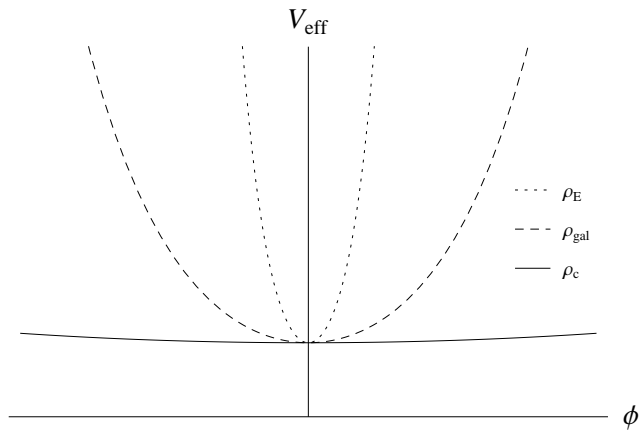}
\caption{Qualitative behaviour of the interacting potential (\ref{042}) for different values of $\rho$. On Earth ($\rho_{\rm E}$) the scalar field is highly massive, while for galactic densities ($\rho_{\rm gal}$) the mass is lower. At cosmological scales ($\rho_c$) the mass is so low that the potential effectively becomes a cosmological constant.}
\label{fig:chameleon}
\end{figure}

For the particular model we will consider here, we will take a scalar field with vanishing self-interaction, i.e.~with $V(\phi)=0$.
Without a coupling to the matter sector such massless scalar field could never accelerate the late time expansion of the universe and thus it would never constitute a viable dark energy model.
We will however assume an interacting term of the general form
\begin{align}
	V_{\rm eff} = f(\rho,\phi) = \Lambda \cosh \left[ \beta_s \frac{\phi}{M_{\rm P}} \left(\frac{\rho}{M_{\rm P}^4}\right)^\alpha \right] \,,
	\label{042}
\end{align}
where $\beta_s$ and $\alpha$ are two dimensionless parameters, $\Lambda$ is the cosmological constant and $M_{\rm P}=1/\kappa=1/\sqrt{8\pi G}$ is the reduced Planck mass.
For any fixed value of the matter energy density $\rho$, the mass of the scalar field is given by
\begin{align}
	m_\phi = \beta_s\frac{\sqrt{\Lambda}}{M_{\rm P}}\left(\frac{\rho}{M_{\rm P}^4}\right)^\alpha \,.
\end{align}
To drive the late time accelerated expansion of the universe, the inverse of this mass, i.e.~the Compton wavelength $\lambda_\phi$, should be of the order of the Hubble length $1/H_0$ at cosmological scales, where the energy density is roughly $\rho_c\simeq 10^{-29} {\rm g/cm^3}$.
In this manner the effective potential (\ref{042}) reduces to nothing but the standard cosmological constant for sufficiently large distances and thus it provides an accelerated expansion with the simplest possible mechanism.
On the other hand if $\rho$ is sufficiently high around laboratories on Earth, then the interaction mediated by the scalar field will be efficiently suppressed because of its large mass.
This situation is qualitatively depicted in Fig.~\ref{fig:chameleon} where it is shown how the potential (\ref{042}) changes for different values of $\rho$.

Imposing now the condition $\lambda_\phi = 1/H_0$ at $\rho_c$, one finds
\begin{align}
	\beta_s = H_0 \frac{M_{\rm P}}{\sqrt{\Lambda}} \left(\frac{M_{\rm P}}{\rho_c}\right)^\alpha \simeq 10^{10+122 \alpha} \,,
\end{align}
which given any positive value of $\alpha$ represents an anomalously large number for the coupling constant $\beta_s$.
Nevertheless, leaving this problem aside, one can now look at the values the mass $m_\phi$ takes on the Earth and within a galaxy.
Using the order of magnitude values $\rho_{\rm E}\simeq 10 {\rm g/cm^3}$, $\rho_{\rm atm}\simeq 10^{-3} {\rm g/cm^3}$ and $\rho_{\rm gal}\simeq 10^{-24} {\rm g/cm^3}$ respectively for the energy density inside the Earth, in our atmosphere and in the galaxy, one finds
\begin{align}
	\lambda_\phi^E \simeq \frac{10^{-30 \alpha}}{H_0} \,,\qquad \lambda_\phi^{\rm atm} \simeq \frac{10^{-26 \alpha}}{H_0} \,,\qquad \lambda_\phi^{\rm gal} \simeq \frac{10^{-5 \alpha}}{H_0} \,.
\end{align}
As an example we consider the values $\alpha=1$ and $\alpha=2$ which yield
\begin{align}
	\lambda_\phi^E &\simeq 10^{-4} \,{\rm m} \,, & \lambda_\phi^{\rm atm} &\simeq 1 \,{\rm m} \,, & \lambda_\phi^{\rm gal} &\simeq 10^{21} \,{\rm m} \simeq 100\, {\rm kpc} \,, & (\alpha &=1) \\
	\lambda_\phi^E &\simeq 10^{-34} \,{\rm m} \,, & \lambda_\phi^{\rm atm} &\simeq 10^{-26} \,{\rm m} \,, & \lambda_\phi^{\rm gal} &\simeq 10^{16} \,{\rm m} \simeq 1\,{\rm pc} & (\alpha &=2) \,.
\end{align}
In both cases the interactions mediated by $\phi$ are short-range on the Earth and long-range on galactic scales.
In the case $\alpha=2$ the mass on the Earth is so high that any effect of the scalar field is practically undetectable.
Nevertheless in both cases at galactic scales the scalar field characterises a long-range force which might give rise to the interesting phenomenology related to the anomalous rotations of galaxies discussed in the previous subsection.
The same fifth-force could also produce testable effects at the level of clustering scales which are easy to analyse within linear cosmological perturbations theory.
From these considerations we obtain that within a single scalar field model defined by the interaction (\ref{042}) one could in principle explain dark energy, dark matter and automatically pass all Solar System constraints.

These brief calculations show the potential of the approach developed in this paper as a powerful tool to construct efficient screening mechanisms which can also provide desirable effects at both galactic and cosmological scales.
We will now turn to the study of cosmological applications of the framework of Sec.~\ref{sec:algebraic_formulation} and leave further analyses at Solar System and galactic distances, as well as at the linear perturbation level, for future investigations.

\section{Cosmology}
\label{sec:cosmology}

In this section we will consider the evolution of the Universe as a whole using models based on the interacting formulation developed in Sec.~\ref{sec:algebraic_formulation} and deriving suitable equations for their dynamics.
The fluid matter sector will now describe dark matter and thus the interacting term between the scalar field and the fluid variables must be considered as an effective coupling between dark energy and dark matter.
Note that due to the results of Sec.~\ref{sec:small_scales} we could also consider a viable coupling to luminous matter if an efficient screening acts at Solar System distances. Nevertheless we will only analyse a coupling to dark matter since this dark component dominates over baryonic matter in the cosmological matter sector.

\subsection{Cosmological equations}

We start considering a Friedmann-Robertson-Walker (FRW) line element as required by the cosmological principle. The metric will then be determined by
\begin{align}
ds^2 = -dt^2 + a(t)^2 \left(\frac{dr^2}{1-k r^2} + r^2 d\Omega^2\right) \,,
\end{align}
where $a(t)$ is the cosmological scale factor, $k=-1,0,1$ according to the spatial openness, flatness or closeness and $d\Omega^2$ is the two-dimensional line element of a sphere. We will also assume that all the dynamical quantities are homogeneous, i.e.~they depend only on $t$. In particular we will have that $\phi$, $\rho$, $n$, $s$ will be functions of $t$ only. Moreover taking into account coomoving coordinates the perfect fluid 4-velocity becomes simply $U^\mu=(-1,0,0,0)$.

The cosmological dynamics is determined by Eqs.~(\ref{019}), (\ref{016}) and (\ref{018}).
However in a cosmological setting we will always have that Eqs.~(\ref{019}) yields
\begin{align}
  \dot n +3H n =0 \quad\mbox{and}\quad \dot s=0 \,,
  \label{eqn:bg0}
\end{align}
where the over-dot denotes differentiation with respect to $t$ and $H=\dot a/a$. These equations tells us that the entropy density per particle is conserved through the Universe evolution, while the particle density decays according to
\begin{align}
n\propto a^{-3} \,,
\end{align}
which is expected from geometric considerations. The only two dynamical quantities that determine the evolution of the Universe are thus $a$ and $\phi$ whose behavior is controlled by Eqs.~(\ref{016}) and (\ref{018}). In a FRW cosmology those equations yield three independent evolution equations, namely the two Friedmann equations
\begin{align}
  3\frac{k}{a^2}+3H^2 &= \kappa^2\left(\rho +\frac{1}{2}\dot\phi^2 +V +\rho_{\rm int} \right) \,,
  \label{eqn:bg1}\\
  \frac{k}{a^2}+2\dot H+3H^2 &=-\kappa^2\left(p+\frac{1}{2}\dot\phi^2 -V +p_{\rm int}\right) \,,
  \label{eqn:bg2}
\end{align}
and the scalar field equation
\begin{align}
\ddot\phi +3H\dot\phi +\frac{\partial V}{\partial\phi} +\frac{\partial\rho_{\rm int}}{\partial\phi} = 0 \,.
\label{025}
\end{align}

Since at the cosmological scales all the dynamical fields are effectively homogeneous, the right hand side of Eq.~(\ref{020}) vanishes. More explicitly for every dynamical quantity $\chi$ we have $\chi_{,\mu}=-\dot\chi\, U_\mu$ and thus $\chi_{,\mu}h^{\mu\nu}=0$. It follows that the matter equation of motion are not modified at large cosmological scales
\begin{align}
\dot\rho +3H\left(\rho+p\right) = 0 \,,
\label{050}
\end{align}
and, assuming $p=w\rho$, the energy density decays exactly as in a non interacting universe
\begin{align}
\rho\propto a^{-3(1+w)} \,.
\end{align}
This is in agreement with the conservations equations (\ref{036}) since in a homogeneous and isotropic cosmological framework the transverse part of those equations vanishes; see below for a derivation using the cosmological equations.

At late times $a\gg 1$ the cosmological dynamics will be completely determined by the scalar field $\phi$ since both $n$ and $\rho$ decay and $\rho_{\rm int}$ and $p_{\rm int}$ will depend on $\phi$ only. If in our model we have that $p_{\rm int}=0$ at late times, then $\rho_{\rm int}$ just describes further dust (dark matter) and, provided $\rho_\phi\gg\rho_{\rm int}$ at late times, the universe evolution will be determined by the form of the potential $V(\phi)$. Choosing a suitable quintessence potential will thus gives rise to a late time cosmic speed-up as required by observations. 

\subsection{The conservation equation}

As it is not trivial to see that the standard matter conservation equation indeed holds, we will show this explicitly in the following. We consider the derivative of~(\ref{eqn:bg1}) and the combination~(\ref{eqn:bg1}) minus~(\ref{eqn:bg2}). Respectively, they give
\begin{align}
  -3H\Bigl(2\frac{k}{a^2}-2\dot{H}\Bigr) &= \kappa^2(\dot{\rho}+\dot{\phi}\ddot{\phi}+V'\dot{\phi}+\dot{\rho}_{\rm int}) \\
  2\frac{k}{a^2}-2\dot{H} &= \kappa^2 (\rho + p + \dot{\phi}^2 + \rho_{\rm int} + p_{\rm int}).
\end{align}
Therefore we get
\begin{align}
  \dot{\rho}+\dot{\phi}\ddot{\phi}+V'\dot{\phi}+\dot{\rho}_{\rm int} = -3H(\rho + p + \dot{\phi}^2 + \rho_{\rm int} + p_{\rm int}).
\end{align}
A few algebraic manipulations give the equation
\begin{align}
  \dot{\rho} + 3H(\rho + p) +\dot{\phi}\Bigl[\ddot{\phi}+3H\dot{\phi}+V'\Bigr]
  + \dot{\rho}_{\rm int} + 3H(\rho_{\rm int} + p_{\rm int}) = 0.
  \label{eqn:c4}
\end{align}
Using (\ref{024}), the last terms involving $\rho_{\rm int}$ and $p_{\rm int}$ can be expanded as
\begin{align}
  \frac{d}{dt}\rho_{\rm int}+ 3H(\rho_{\rm int} + p_{\rm int}) &= \frac{\partial \rho_{\rm int}}{\partial n}\dot{n} +  \frac{\partial \rho_{\rm int}}{\partial s}\dot{s} + \frac{\partial\rho_{\rm int}}{\partial \phi}\dot{\phi} + 3H n\frac{\partial\rho_{\rm int}}{\partial n} \\
&=\frac{\partial \rho_{\rm int}}{\partial n}\left(\dot{n}+3H n\right) +  \frac{\partial \rho_{\rm int}}{\partial s}\dot{s} + \frac{\partial\rho_{\rm int}}{\partial \phi}\dot{\phi} 
\end{align}
The first and second term are zero due to~(\ref{eqn:bg0}). Now we can rewrite~(\ref{eqn:c4}) in the following way
\begin{align}
  \dot{\rho} + 3H(\rho + p) +\dot{\phi}\Bigl[\ddot{\phi}+3H\dot{\phi}+
    V'+\frac{\partial \rho_{\rm int}}{\partial \phi}\Bigr]
   = 0.
  \label{eqn:c6}
\end{align}
The term proportional to $\dot{\phi}$ is zero because of the equation of motion of the scalar field (\ref{025}). Therefore, we have proved that 
\begin{align}
  \dot{\rho} + 3H(\rho + p) = 0 \,,
  \label{eqn:c8}
\end{align}
in agreement with Eqs.~(\ref{050}).

\subsection{Relation with standard interacting models}

In the following we will clarify the relationship between the variational approach used here and the frequently used approach of introducing couplings at the level of the conservation equations. Using the notation $\tilde{\rho} = \rho + \rho_{\rm int}$ and $\tilde{p} = p + p_{\rm int}$ one can rewrite Eq.~(\ref{eqn:c4}) in the form
\begin{align}
  \dot{\tilde{\rho}} + 3H(\tilde{\rho} + \tilde{p}) +\dot{\phi}\Bigl[\ddot{\phi}+3H\dot{\phi}+V'\Bigr] = 0,
  \label{eqn:c4aa}
\end{align}
which is the sum of two conservation equations, that of the fluid with $\tilde{\rho}$, $\tilde{p}$, and the conservation equation of the scalar field. 
Let us now recall the equation of motion of the scalar field~(\ref{025}) which we write
\begin{align}
  \ddot\phi +3H\dot\phi +\frac{\partial V}{\partial\phi} = \frac{Q}{\dot{\phi}} 
  \quad\mbox{with}\quad Q = -\frac{\partial\rho_{\rm int}}{\partial\phi} \dot{\phi},
\end{align}
so that it takes the standard form with coupling $Q$, see also Eqs.~(\ref{eq:standard_coupling}). Consequently, Eq.~(\ref{eqn:c4aa}) implies a coupled equation for the matter
\begin{align}
  \dot{\tilde{\rho}} + 3H(\tilde{\rho} + \tilde{p}) = -Q.
\end{align}
Note that these equations can equivalently be derived from the general covariant equations (\ref{eq:scalar_conserv_cov_tilde}) and (\ref{eq:matter_cov_tilde}).

We are now in a position to link the two different approaches to coupled cosmological models. 
Our approach is equivalent to standard coupled models provided we identify $\tilde{\rho}$ and $\tilde{p}$ with the usual matter variables $\rho$ and $p$ in the standard approach. In other words, the standard approach uses an effective description of the matter sources present. The quantity $\tilde{\rho}$, for instance, contains a scalar field dependence and therefore the standard approach does not allow the consistent separation into matter and interaction parts without additional ad-hoc assumptions.
This is not an issue solely concerned within scalar filed models, but it equally arises in any dark energy model interacting with dark matter, as recently pointed out in \cite{Tamanini:2015iia}.

At least for scalar fields, we can conclude that every standard coupled model can be derived from the Lagrangian formulation presented here. However, it is also clear that the correct identification of matter is required to make this approach consistent.

\subsection{Dynamical system techniques}

The subsequent analysis of the background dynamics relies on the use of dynamical systems techniques. In order to be self-contained, we briefly introduce the most important concepts tailored to three-dimensional systems. A three-dimensional autonomous system of differential equations, also called a dynamical system, is given by the three equations
\begin{align}
  x' = f_1(x,y,z), \quad y' = f_2(x,y,z), \quad z' = f_3(x,y,z)\,,
  \label{eq:sys1}
\end{align}
where the prime denotes differentiation with respect to a suitable time parameter. The three functions $f_i(x,y,z)$ do not explicitly depend on the time parameter, they only depend on the dynamical variables $x,y,z$.
In technical terms this means we are dealing with {\it autonomous systems} only.

A {\it fixed point}, {\it critical point} or {\it stationary point} of the system~(\ref{eq:sys1}) is a point with coordinates $(x_0,y_0,z_0)$ such that $f_i(x_0,y_0,z_0) = 0$ for $i=1,2,3$. This corresponds to a stationary point of a mechanical system where the potential has an extremal point and a fictitious particle would remain at rest. In analogy, at the point $(x_0,y_0,z_0)$ the system is at rest since all evolution equations are identically satisfied; in principle the system could remain in this state indefinitely. 

Linear stability theory clarifies whether a stationary point is stable or unstable with respect to small perturbations away from that point. The basic idea behind this approach is to Taylor expand the three functions $f_i(x,y,z)$ around the fixed point $(x_0,y_0,z_0)$. Since the functions vanish at the fixed point, the first non-trivial terms in the Taylor series will involve the first partial derivatives of the functions. Let us denote $x_j=(x,y,z)$ for $j=1,2,3$. It is therefore natural to define the matrix of first derivatives
\begin{align}
  J = \frac{\partial f_i}{\partial x_j}, \quad i,j=1,2,3\,,
\end{align}
which is the familiar Jacobian matrix of vector calculus (not to be confused with the current $J^\mu$). The information on stability is contained in the eigenvalues of this matrix $J$ evaluated at the critical point $(x_0,y_0,z_0)$. Since $J$ is a $3 \times 3$ matrix, it will have three, not necessarily distinct, eigenvalues. If all eigenvalues of $J$ are positive, we speak of an unstable point since all perturbations would grow exponentially. On the other hand, if all eigenvalues are negative this point would be regarded stable. If at some fixed point the matrix $J$ contains positive and negative eigenvalues, then one speaks of a saddle point. 

One could also encounter a pair of complex conjugate eigenvalues and one additional real eigenvalue. In this case, stability and instability will depend on the signs of the real parts of the eigenvalues. Linear stability theory breaks down when at least one of the eigenvalues has a zero real part. In this case one has to use techniques beyond linear stability theory to understand the dynamics of the system near this fixed point, see for instance~\cite{Boehmer:2014vea}. However, these techniques will not be required in the following as linear stability theory turns out to be sufficient.   

\subsection{Cosmological dynamics}

In this section we will study the dynamics of a universe described by Eqs.~(\ref{eqn:bg1})-(\ref{025}) employing dynamical systems methods to determine its complete dynamical properties. Similar analyses for dark energy models coupled directly through the cosmological equations, i.e.~considering Eqs.~(\ref{eq:standard_coupling}) for different phenomenological choices of $Q$, have already been performed in several works; see e.g.~\cite{Amendola:1999er,Boehmer:2008av,Boehmer:2011tp}.
We introduce the following dimensionless variables
\begin{align}
\sigma^2=\frac{\kappa^2\rho}{3H^2}\,, \quad x^2=\frac{\kappa^2\dot\phi^2}{6H^2}\,,\quad
y^2=\frac{\kappa^2V}{3H^2}\,,\quad
z=\frac{\kappa^2\rho_{\rm int}}{3H^2}\,,
\label{026}
\end{align}
which straightforwardly generalise the normalised variables usually employed to analyse the dynamics of quintessence \cite{Copeland:1997et}.
In general, since $\rho_{\rm int}$ is a function of $n$ and $\phi$ ($s$ is constant because of (\ref{eqn:bg0}) and will not be considered in what follows) which in turn can be seen as functions of $\rho$ and $V$ respectively, we can always write it as a combination of the variables $\sigma$ and $y$. However, unless the function $\rho_{\rm int}$ is specifically choosen not to do so, this will require the introduction of another variable such as $\tilde{z}=H/(H_0+H)$ with $H_0$ a constant. To directly consider a fourth variable $z$ as defined in (\ref{026}) simplifies the analysis since it allows us to reduce the Friedmann equation (\ref{eqn:bg1}) to the constraint
\begin{align}
1=\sigma^2+x^2+y^2+z \,,
\label{eqn:bg1a}
\end{align}
which will permit us to replace $\sigma$ in terms of the other variables. Moreover we consider only $y>0$ since we can assume $V>0$. Alternatively we will see that the system will be invariant under the changes $y\mapsto-y$, meaning that its dynamics will be specular over the $xy$ planes.

In what follows we will assume that $p=w\rho$ with $w$ a constant, called the matter equation of state parameter, physically constrained to be between $0$ and $1/3$.
For (dark) matter one has $w=0$, however we will keep $w$ as a general constant for the moment.
We will also assume the potential $V$ to have an exponential form of the kind
\begin{align}
V(\phi) = V_0\, e^{-\lambda\kappa\phi} \,,
\label{055}
\end{align}
with $V_0$ a constant and $\lambda$ a dimensionless parameter. The exponential form for $V$ is the only one which consents to close the autonomous system of equations without introducing another variable.
This happens because $V'\propto V$ and thus derivatives of the potential can be related to $y$.
For different potentials instead a further variable must be added in order to take into account the derivatives of the scalar field potential \cite{Ng:2001hs}.

With these assumptions the cosmological field equations (\ref{eqn:bg1})-(\ref{025}) can be rewritten as
\begin{align}
  x' &= -\frac{1}{2} \left(3x \left((w+1) y^2+w z-w+1\right) + 3 (w-1) x^3-\sqrt{6} \lambda  y^2\right) + x A -B \label{071}\\
  y' &= -\frac{1}{2} y \left( 3 (w-1) x^2 + 3 \left((w+1) y^2+w z - w-1\right)+\sqrt{6} \lambda  x\right) + y A \label{072}\\
  z' &= 2 A (z-1)+2 B x-3 z \left((w-1) x^2 + (w+1) y^2 + w (z-1)\right)\label{073}.
\end{align}
where a prime denote differentiation with respect to $d\eta=H dt$ and we have defined
\begin{align}
A=\frac{\kappa ^2 p_{\rm int}}{2 H^2} \quad\mbox{and}\quad B=\frac{\kappa}{\sqrt{6} H^2}\frac{\partial\rho_{\rm int}}{\partial\phi} \,.
\end{align}
In order to close the system we must specify the function $\rho_{\rm int}$ and then compute $A$ and $B$ using (\ref{024}).
If $\rho_{\rm int}$ is chosen accordingly the quantities $A$ and $B$ become functions of $x$, $y$, $z$ and the system results to be close.
If instead $A$ and $B$ cannot be written as functions of $x$, $y$, $z$ then another variable has to be added in the dynamics and the dimensions of the system increase.

The following choices do not increase the dimensions of the system further:
\begin{table}[!ht]
\begin{tabular}{c|c|c|c|c}
\mbox{} & $\rho_{\rm int}$ & $p_{\rm int}$ & $A$ & $B$ \\
\hline
Model I & $\gamma\,\rho^\alpha\exp(-\beta\kappa\phi)$ & $[\alpha(w+1)-1]\rho_{\rm int}$ & $\frac{3}{2}[\alpha(w+1)-1]z$ & $-\beta\sqrt{\frac{3}{2}}z$ \\
Model II &
$\gamma\kappa\phi\rho$ & $w\rho_{\rm int}$ & $\frac{3}{2}wz$ & $\gamma\sqrt{\frac{3}{2}}\left(1-x^2-y^2-z\right)$
\end{tabular}
\end{table}
where $\alpha$, $\beta$ and $\gamma$ are dimensionless parameters.
These choices are the simplest ones found by the authors which do not require a dynamical systems of more than three dimensions, i.e.~for which $A$ and $B$ can be written as functions of $x$, $y$, $z$.
Mathematically they are simple to analyze and physically they are sufficiently complex to allow us to investigate the new and rich phenomenology of the scalar-fluid coupling considered in this work.

These two models are also characterised by a linear equation of state $p_{\rm int}=w_{\rm int}\rho_{\rm int}$ with $w_{\rm int}$ being constant and thus they represent the simplest three-dimensional models one can think of. However, by considering other forms of $\rho_{\rm int}$, one can construct models where the equation of state $w_{\rm int}$ itself becomes dependent on the variables $x,y,z$.
One such choice is $\rho_{\rm int} = \gamma V \exp(\rho/\rho_0)$ where $\rho_0$ is some constant characteristic density. This gives $p_{\rm int} = [(1+w)\rho/\rho_0 - 1] \rho_{\rm int}$ which means $w_{\rm int} = (1+w)\rho/\rho_0 - 1$. Recall that $\rho$ is related to $\sigma^2$ which satisfies (\ref{eqn:bg1a}) and thus introduces the dependence on $x,y,z$.

Unfortunately, such a model would increase the dimensions of the corresponding dynamical system since $w_{\rm int}$ cannot be written solely in terms of the variables $x$, $y$, $z$. A coupling where this can be achieved corresponds for example to the rather complicated choice $\rho_{\rm int}=\gamma V/\ln(\rho/\rho_0)$ with $V(\phi)$ the potential (\ref{055}). In this case one obtains $w_{\rm int}=-1-(w+1)z/(\gamma y^2)$ and both $A$ and $B$ can be written as functions of $x$, $y$, $z$ without introducing other dynamical variables.

In what follows we will consider only the two simplest choices corresponding to a linear interacting equation of state and leave analyses of the more complicated models for future studies.
We will label them as Model I and II as outlined by the table above.

\subsubsection{Model I}

In this subsection we analyse Model I of the dynamical system~(\ref{071})-(\ref{073}). Now the two Friedmann equations~(\ref{eqn:bg1}) and (\ref{eqn:bg2}) can be written to give the acceleration equation
\begin{align}
\frac{\dot{H}}{H^2}=\frac{3}{2}[- (1+w) + (w-1)x^2 + (1+w)y^2 - (1+w) (\alpha-1) z]\label{075}
\end{align}
which can be solved for $a$ at any fixed point $(x_*,y_*,z_*)$ of the phase space to give
\begin{align}
a \propto (t-t_0)^{2/3[ (1+w) - (w-1)x_*^2 - (1+w)y_*^2 + (1+w) (\alpha-1) z_*]}\label{076}
\end{align}
where $t_0$ is a constant of integration. This represents a power law solution, where the scale factor $a$ evolves as a power of cosmological time $t$. If the denominator of~(\ref{076}) vanishes, this forces $H$ to be constant, which corresponds to the universe undergoing a de Sitter expansion. If we define the effective energy density and pressure of the system to be
\begin{align}
\rho_{\rm eff}&=\rho+\frac{1}{2}\dot{\phi}^2+V +\rho_{\rm int} \\
p_{\rm eff}&=p+\frac{1}{2}\dot{\phi}^2-V+p_{\rm int}
\end{align}
we can then define an effective equation of state paramter $w_{\rm eff}$, which can be expressed in terms of the dimensionless variables~(\ref{026}) 
\begin{align}
  w_{\rm eff} &= \frac{p_{\rm eff}}{\rho_{\rm eff}} =
  w - (w-1)x^2 - (1+w)y^2 + (1+w) (\alpha-1) z.
\end{align}
These definitions allow us to write~(\ref{076}) in the simple form
\begin{align}
a \propto (t-t_0)^{2/[3(1+w_{\rm eff})]}.\label{078}
\end{align}
The energy density and pressure of the scalar field are given by
\begin{align}
\rho_\phi&=\frac{1}{2}\dot{\phi}^2+V  \\
p_\phi&= \frac{1}{2}\dot{\phi}^2-V
\end{align}
and we can also define the equation of state of the scalar field at any point in our phase space
\begin{align}
  w_\phi &= \frac{p_\phi}{\rho_\phi} = \frac{x^2 - y^2}{x^2 + y^2}.
\end{align}

Now that we have specified the model, we can find the critical/fixed points of the autonomous system~(\ref{071})--(\ref{073}). These are defined to be the points $(x,y,z)$ of phase space which satisfy
\begin{align}
x'=0, \quad y'=0, \quad z'=0
\end{align}
If the system is at one of these points, there is no dynamical evolution and the universe evolves according to~(\ref{078}). In order to exist, the critical points must lie in the three dimensional phase space, which are all points $(x,y,z)$ such that
\begin{align}
x^2+y^2+z\leq 1.
\end{align}
This follows from the assumption $\sigma^2>0$, i.e.~$\rho>0$, and the Friedmann constraint (\ref{eqn:bg1a}).
Stability of critical points is determined by linearising the autonomous system around the critical point under consideration, which leads to analysing the eigenvalues of the Jacobian matrix
\begin{align}
\mathcal{M}_{ij} = \frac{\partial f_i}{\partial x^j} 
\end{align}
evaluated at the critical point, where we have compactly written the dynamical system~(\ref{071})-(\ref{073}) as $x'=f_x(x,y,z)$, $y'=f_y(x,y,z)$ and $z'=f_z(x,y,z)$. If all three eigenvalues of $\mathcal{M}$ have negative real part the critical point is stable, if all three have positive real part the point is unstable, and if some eigenvalues have different signs the point is a saddle.

There are up to eight critical points of the system, depending on the values of the parameters $\alpha$, $\beta$, $w$ and $\lambda$. They are outlined in Tab.~\ref{tab:m1_gen}.

\begin{table}[!ht]
\begin{tabular}{|c|c|c|c|}
  \hline
  Point & $x$ & $y$ & $z$ \\
  \hline \hline 
  O & 0 & 0 & 0 \\
  \hline
  $A_{\pm}$ & $\pm 1$ & 0 & 0 \\
  \hline
  $B$ & $\sqrt{\frac{3}{2}} \frac{(1+w)}{\lambda}$ & 
  $\sqrt{\frac{3}{2}} \frac{\sqrt{(1+w)(1-w)}}{\lambda}$ & 0 \\
  \hline
  $C$ & $\frac{\lambda}{\sqrt{6}}$ & $\sqrt{1-\frac{\lambda^2}{6}}$ &
  $0$ \\
  \hline
  $D$ & $-\sqrt{\frac{2}{3}} \frac{\beta}{\alpha(w+1)-2}$ &
  0 & $1 - \frac{2\beta^2}{3(\alpha+\alpha w-2)^2}$ \\
  \hline
  $E$ & $\sqrt{\frac{3}{2}} \frac{(1+w)(1-\alpha)}{\beta}$ &
  0 & $\frac{3}{2} \frac{(1-\alpha)(1+w)(1-w)}{\beta^2}$ \\
  \hline
  $F$ & $\sqrt{\frac{3}{2}} \frac{(1+w)\alpha}{\lambda-\beta}$ &
  $\frac{\sqrt{6(w+1)\alpha-3(w+1)^2\alpha^2+2\beta(\beta-\lambda)}}{\sqrt{2}|\lambda-\beta|}$ &
  $\frac{\lambda(\lambda-\beta)-3(1+w)\alpha}{(\beta-\lambda)^2}$ \\
  \hline
\end{tabular}
\caption{Critical points of Model I.}
\label{tab:m1_gen}
\end{table}

We have checked explicitly that there are no critical points at infinity. We did this by using the compact variable $\zeta = \arctan(z)$ and it turns out that there are no critical points where $\zeta = \pm \pi/2$.\\
$ $

{\bf The case $\alpha=1$}.\\
The dynamics of the system depends on the four constants $\alpha$, $\beta$, $\lambda$ and $w$. In order to simplify the analysis we will first consider the natural choice $\alpha=1$. This particular choice makes the critical point $E$ coincide with the origin $O$. We will also only analyse the dynamics in a matter dominated universe with $w=0$, the dynamics of the system for other values of $w$ are qualitatively the same and not relevant for models interacting with (cold) dark matter.
We are thus left with the critical points listed in Tab.~\ref{tab:m1_a1}.

\begin{table}[!ht]
\begin{tabular}{|c|c|c|c|}
  \hline
  Point & $x$ & $y$ & $z$ \\
  \hline \hline 
  O & 0 & 0 & 0 \\
  \hline
  $A_{\pm}$ & $\pm 1$ & 0 & 0 \\
  \hline
  $B$ & $\sqrt{\frac{3}{2}} \frac{1}{\lambda}$ & 
  $\sqrt{\frac{3}{2}} \frac{1}{\lambda}$ & 0 \\
  \hline
  $C$ & $\frac{\lambda}{\sqrt{6}}$ & $\sqrt{1-\frac{\lambda^2}{6}}$ &
  $0$ \\
  \hline
  $D$ & $\sqrt{\frac{2}{3}}\beta$ &
  0 & $1 - \frac{2\beta^2}{3}$ \\
  \hline
  $F$ & $\sqrt{\frac{3}{2}} \frac{1}{\lambda-\beta}$ &
  $\frac{\sqrt{3+2\beta(\beta-\lambda)}}{\sqrt{2}|\lambda-\beta|}$ &
  $\frac{\lambda(\lambda-\beta)-3}{(\beta-\lambda)^2}$ \\
  \hline
\end{tabular}
\caption{Critical points of Model I with $w=0$ and $\alpha=1$.}
\label{tab:m1_a1}
\end{table}

\begin{figure}[!ht]
\includegraphics[width=0.48\textwidth]{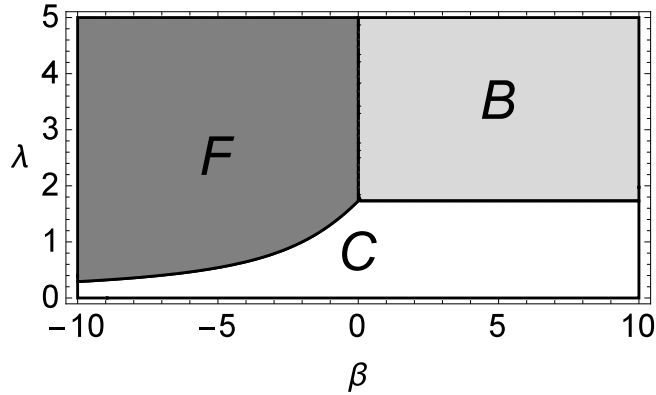}
\caption{$\beta$-$\lambda$ parameter space indicating the region where the critical points $B,C$ and $F$ are stable.}
\label{F stability}
\end{figure}

Properties of these critical points, including existence and stability can be found in table~\ref{tab01}.  There are potentially up to seven critical points depending on the values of $\lambda$ and $\beta$:
\begin{itemize}
\item {\it Point $O$}. The origin of the phase space exists for all values of $\lambda$ and $\beta$ and corresponds to a matter dominated universe. In this case $w_{\rm eff}=0$ (or more generally $w_{\rm eff}=w$), so there is no acceleration at this point. This point is always a saddle point.
\item {\it Point $A_{\pm}$}. These two points are dominated by the scalar field kinetic energy, with the effective equation of state reducing to that of a stiff fluid $w_{\rm eff}=1$. Thus no acceleration is present at this point. These points are either unstable or saddle points depending on whether the absolute values of $\lambda$ and $\beta$ are less than $\sqrt{6}$ and $3/\sqrt{6}$ respectively. 
\item {\it Point B.} This point exists for all values of $\lambda$ and $\beta$. It corresponds to the usual scaling solution where the effective equation of state matches the matter equation of state, yet the scalar field energy density does not vanish. Hence the universe behaves as if it were completely matter dominated, yet the energy density of both the matter field and scalar field do not vanish. This point is only stable when $\beta>0$. 
\item {\it Point C.} This point corresponds to a universe completely dominated by a scalar field. It exists only for $\lambda^2<6$. It is stable when $\lambda^2<3$ and $\lambda \beta>\lambda^2-3$, and a saddle node otherwise. This point is the usual cosmological accelerated expansion driven by a sufficiently flat scalar potential: $\lambda^2<2$. 
\item {\it Point D.} This point exists for all values of $\lambda$ and $\beta $ and is specific of interacting quintessence models∫. The effective equation of state parameter is always positive, so this point never corresponds to an accelerating universe. This point is either unstable or a saddle point depending on the values of $\beta$ and $\lambda$. In the limit $\beta \rightarrow 0$, this point merges with the origin. 
\item {\it Point F.} This point exists only when $\beta(\beta-\lambda)>-3/2$. The energy density of the scalar field and interaction components are non-zero, and the energy density of matter vanishes. This point can correspond to an accelerating solution when the parameters lie in the range $-2<\frac{\lambda}{\beta}<1$, and corresponds to either a stable spiral or a saddle node. The region in parameter space where this point is stable is indicated by region $F$ in Fig.~\ref{F stability}. Thus we can conclude there are points in parameter space where this point describes a late time cosmological accelerating attractor solution. 
\end{itemize}

\begin{table}[ht]
\begin{tabular}{|c|c|c|c|c|}
\hline
\mbox{Point} & Existence & $w_{\rm eff}$ & Acceleration & Stability \\
\hline \hline
$O$ & $\forall \lambda,\beta$ & $w$ & No & Saddle node\\
\hline
$A_{-}$ & $\forall \lambda,\beta$ & 1 & No & Unstable node: $\beta> \frac{-3}{\sqrt{6}}, \lambda>-\sqrt{6}$ \\
& & & & Saddle node: otherwise \\
\hline
$A_{+}$ & $\forall \lambda,\beta$& 1 & No & Unstable node: $\beta< \frac{3}{\sqrt{6}}, \lambda<\sqrt{6}$ \\
& & & & Saddle node: otherwise \\
\hline
$B$ & $\lambda^2>3$ & $w$ & No & Stable node: $3<\lambda^2<24/7\: \& \: \lambda\beta>0$  \\
& & & & Stable spiral: $\lambda^2>24/7\: \& \:\lambda \beta>0$ \\
& & & & Saddle node: $\lambda\beta<0$ 
\\ \hline $C$ & $\lambda^2<6$ & $\frac{\lambda^2-3}{3}$ & $\lambda^2<2$ & Stable node: $\lambda^2<3  \:\&\:\lambda\beta>(\lambda^2-3)$\\
& & & & Saddle node: $\beta<(\lambda^2-3)/\lambda$ \\
\hline
$D$ & $\forall \lambda,\beta$ & $\frac{2\beta^2}{3}$ & No & Unstable node: $\beta^2>3/2 \: \& \: \lambda<(3\beta^2+3)/\beta$ \\
 & & & & Saddle node: otherwise
\\
\hline
$F$ & $3+2\beta(\beta-\lambda)>0$ & $\frac{\beta}{\lambda-\beta}$ &  $-2<\frac{\lambda}{\beta}<1$ & Stable spiral:  region  in Fig.~\ref{F stability}\\
 & & &  & Saddle node: otherwise \\
\hline
\end{tabular}
\caption{Stability of critical points of Model 1 with $w=0$ and $\alpha=1$.} 
\label{tab01}
\end{table}

Analysing the stability of the fixed points, we see that when $\beta>0$ the late time behaviour of the dynamical system is qualitatively the same as the case of the canonical scalar field with no interaction term. Although we now potentially have up to two extra critical points arising, they are either unstable or saddle points, and at late times the solution will either limit towards the scalar field dominated point $C$ or the scaling solution $B$. On the other hand when $\beta<0$ we have new late time behaviour arising. Depending on the parameter values the late time attracting solutions are either the scalar field dominated point $C$, or the point $F$, a point in which the energy density of the matter, scalar field and interaction term is non-zero. This point can represent an accelerating solution for a wide range of parameter values.
Note that Point~$F$ can in principle represent a solution to the cosmic coincidence problem since one can attain an accelerating late time attractor where the energy density of dark energy does not dominate completely (accelerating scaling solution).

\begin{figure}[!ht]
\includegraphics[width=0.48\textwidth]{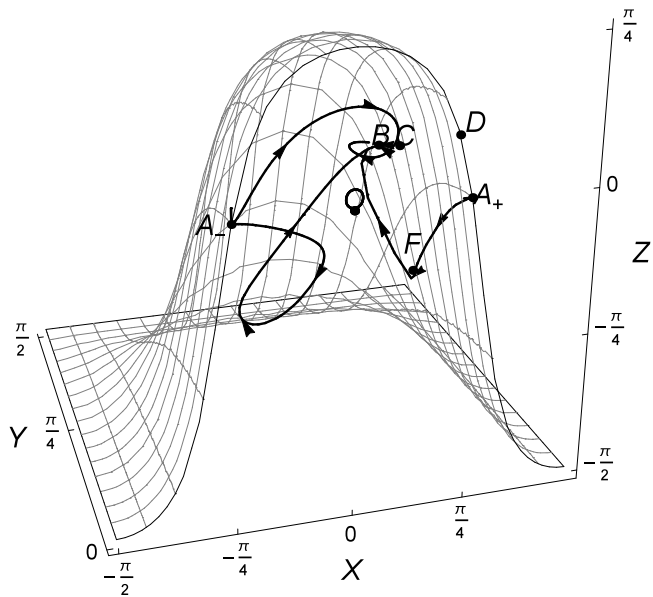}
\caption{Phase space showing trajectories of the dynamical system~(\ref{071})-(\ref{073}) when $\alpha=1,\beta=1,\lambda=2$ and $w=0$. Point $B$ is the global attractor where the universe behaves as though it were completely matter dominated.}\label{alpha1beta1lambda2w0}
\end{figure}

Now the phase space of the dynamical system is the subset of $\mathbb{R}^3$ defined by $-\infty<x<\infty$, $y\geq0$, $-\infty<z\leq1-x^2-y^2$. Hence the phase space is non-compact. In order to plot the phase space, we therefore compactify the system by introducing the following new variables
\begin{align}
X=\arctan x, \quad 
Y=\arctan y, \quad
Z=\arctan z. \label{compactvars} 
\end{align}
The phase space is now compact, with $X,Y,Z$ now lying in the range $-\pi/2<X<\pi/2$, $0\leq Y < \pi/2$ and $-\pi/2< Z< \arctan(1-\tan^2 X-\tan^2 Y)$. This is displayed in Fig.~\ref{alpha1beta1lambda2w0}.  One could have chosen other functions to compactify, for instance, $X = \mathrm{arctanh} \, x$, etc. which leads to the same qualitative results. 

\begin{figure}[!ht]
\includegraphics[width=0.48\textwidth]{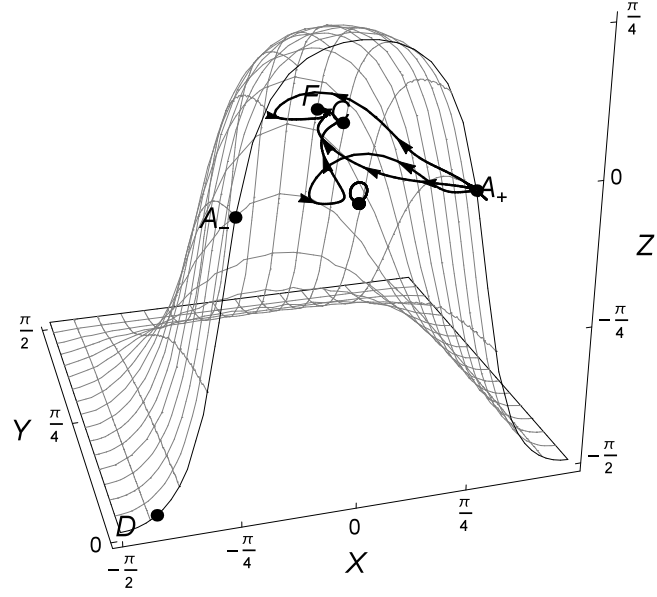}
\caption{Phase space showing trajectories of the dynamical system~(\ref{071})-(\ref{073}) when $\alpha=1,\beta=-5,\lambda=2$ and $w=0$. Point $F$ is the global attractor, and for this choice of parameters it corresponds to a solution with $w_{\rm eff}=-5/6$. }\label{alpha1beta-5lambda2w0}
\end{figure}

We show phase space diagrams with a few trajectories for two distinct cases, when $\beta>0$ and $\beta<0$ in Fig.~\ref{alpha1beta1lambda2w0} and Fig.~\ref{alpha1beta-5lambda2w0} respectively. In Fig.~\ref{alpha1beta1lambda2w0} the trajectories start from either the stiff matter states $A_{\pm}$ and evolve towards point $B$, which serves as the global attractor for this choice of parameters. This dynamics, similar to the case of the canonical scalar field with no interaction term, is well suited to describing the late time phenomenology of our universe as it has trajectories describing a decelerated to accelerated transition, corresponding to the dominance of dark energy over dark matter at late times. However this model suffers from the same issues as the canonical scalar field at early times, since the early time attractors are the points $A_{\pm}$, which have effective equation of state $w_{\rm eff}=1$ which is not physically viable on the classical level.  
In Fig.~\ref{alpha1beta-5lambda2w0} all trajectories again start at either $A_{\pm}$, however they now evolve to the point $F$ which is the global attractor, which for this choice of parameters is a solution with $w_{\rm eff}=-5/6$. Again, for suitable choice of the parameters this model can well describe the late time universe phenomenology, however it breaks down at early times.

{\bf The case $\alpha=3$}.\\
We will now consider the effects on model I of the above dynamical system when we take a different value of $\alpha$. In general when $\alpha\neq1$, the number of fixed points of the system increases. For simplicity, we will not consider $\alpha=2$ as this gives us critical points at infinity, so we will analyse the case $\alpha=3$, again in a matter dominated universe with $w=0$.
Note that values of $\alpha$ different from one correspond to scalar-fluid couplings non-linear in $\rho$.
Such kind of non-linear couplings can only be built using the variational formalism presented in this work, whereas other interacting theories can only couple the scalar filed to $\rho$ linearly, as in the case of Scalar-Tensor theories for example \cite{Amendola:1999qq,Amendola:1999er,Holden:1999hm}.
Depending on the parameter choices, the system now has potentially up to eight fixed points which are displayed in Tab.~\ref{tab:m1_a3}.

\begin{table}[!ht]
\begin{tabular}{|c|c|c|c|}
  \hline
  Point & $x$ & $y$ & $z$ \\
  \hline \hline 
  O & 0 & 0 & 0 \\
  \hline
  $A_{\pm}$ & $\pm 1$ & 0 & 0 \\
  \hline
  $B$ & $\sqrt{\frac{3}{2}} \frac{1}{\lambda}$ & 
  $\sqrt{\frac{3}{2}} \frac{1}{\lambda}$ & 0 \\
  \hline
  $C$ & $\frac{\lambda}{\sqrt{6}}$ & $\sqrt{1-\frac{\lambda^2}{6}}$ &
  $0$ \\
  \hline
  $D$ & $-\sqrt{\frac{2}{3}} \beta$ &
  0 & $1 - \frac{2\beta^2}{3}$ \\
  \hline
  $E$ & $-\frac{\sqrt{6}}{\beta}$ &
  0 & $-\frac{3}{\beta^2}$ \\
  \hline
  $F$ & $\sqrt{\frac{3}{2}} \frac{3}{\lambda-\beta}$ &
  $\frac{\sqrt{-9+2\beta(\beta-\lambda)}}{\sqrt{2}|\lambda-\beta|}$ &
  $\frac{\lambda(\lambda-\beta)-9}{(\beta-\lambda)^2}$ \\
  \hline
\end{tabular}
\caption{Critical points of Model I with $w=0$ and $\alpha=3$.}
\label{tab:m1_a3}
\end{table}

The existence and stability  properties of these critical points, along with the values of $w_{\rm eff}$ are displayed in table~\ref{tab02}.  Qualitatively, the properties of the critical points $O, A_{\pm}, B$ and $C$ remain the same as the $\alpha=1$ case, the only modification being the exact regions of $\beta-\lambda$ parameter space where stability/instability arises are slightly different. However the points $D$, $E$ and $F$ exhibit new behaviour when $\alpha=3$.

\begin{figure}[!ht]
\includegraphics[width=0.75\textwidth]{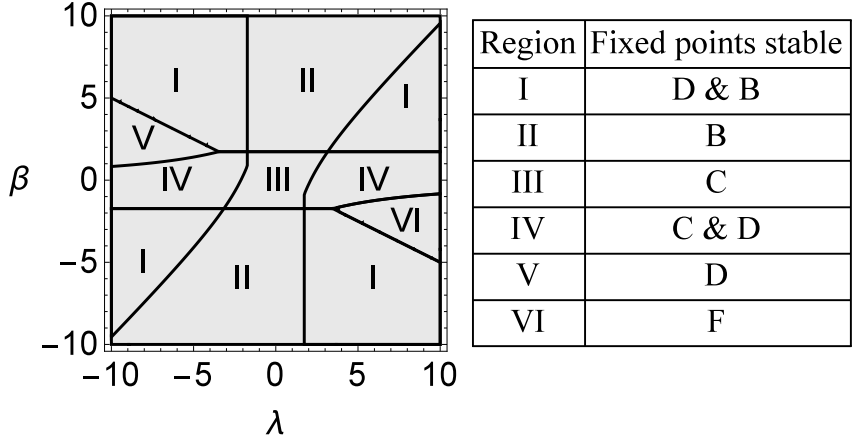}
\caption{$\beta$-$\lambda$ parameter space indicating the different regions where the critical points $B,C,D$ and $F$ are stable.}
\label{alpha3stability}
\end{figure}

\begin{itemize}
\item {\it Point $D$}. This point exists for all values of $\lambda$ and $\beta$. It describes a solution where the kinetic energy of the scalar field and the energy density of the interaction term is non-zero, but the energy density of matter vanishes. Unlike the $\alpha=1$ case, this point can now describe an accelerating solution and is also stable for certain regions of parameter space. If $\beta^2>7/2$ this solution describes an accelerating universe and when $\beta^2>3$ and $2\beta\lambda<2\beta^2-9$, this point is a stable node. Hence whenever the point is stable, it describes an accelerating solution. 
\item{\it Point $E$}. This point only exists when $\beta>3$. The potential energy of the scalar field vanishes, yet the scalar field's kinetic energy, along with the energy density of the matter and interaction fields, are all non-zero. This solution is a scaling solution where the universe behaves as if it were completely matter dominated, yet the universe evolves under the influence of all three present fields. This solution is never a late time attractor, it is either an unstable node or a saddle point.
\item{\it Point $F$}. This point only exists when $2\beta(\beta-\lambda)>9$. As in the $\alpha=1$ case, this point represents a solution with both scalar field and interaction term components, with the energy density of the matter field vanishing. For certain values of parameter space, this solution is a stable spiral and the global attractor, see Fig~\ref{alpha3stability}. The solution is accelerating only if $-2/7<\lambda/\beta<1$. 
\end{itemize}

\begin{table}[!ht]
\begin{tabular}{|c|c|c|c|c|}
\hline
\mbox{Point} & Existence & $w_{\rm eff}$ & Acceleration & Stability \\
\hline \hline
$O$ & $\forall \lambda,\beta$ & $w$ & No & Saddle node\\
\hline
$A_{-}$ & $\forall \lambda,\beta$ & 1 & No & Unstable node: $\beta> \frac{-3}{\sqrt{6}}, \lambda>-\sqrt{6}$ \\
& & & & Saddle node: otherwise \\
\hline
$A_{+}$ & $\forall \lambda,\beta$ & 1 & No & Unstable node: $\beta< \frac{3}{\sqrt{6}}, \lambda<\sqrt{6}$ \\
& & & & Saddle node: otherwise \\
\hline
$B$ & $\lambda^2>3$ & $w$ & No & Stable node: $3<\lambda^2<24/7\: \& \: \beta/\lambda>-2$  \\
& & & & Stable spiral: $\lambda^2>24/7\: \& \:\beta/\lambda>-2$ \\
& & & & Saddle node: $\beta/\lambda<-2$
\\
\hline
$C$ & $\lambda^2<6$ & $\frac{\lambda^2-3}{3}$ & $\lambda^2<1$ & Stable node: $\lambda^2<3  \:\&\:\lambda\beta>(\lambda^2-9)$\\
& & & & Saddle node: $\lambda\beta<(\lambda^2-9)$ or $\lambda^2>3$ \\
\hline
$D$ & $\forall \lambda,\beta$ & $\frac{6-2\beta^2}{3}$ & $\beta^2>7/2$ & Stable node: $\beta^2>3 \: \& \: 2\beta\lambda<(2\beta^2-9)$ \\
 & & & & Unstable node: $\beta^2 <3/2\: \& \: 2\beta\lambda>(2\beta^2-9)$
\\
& & & & Saddle node: otherwise
\\
\hline
$E$ & $\beta^2>3$ & $w$ &  No & Unstable node: $\lambda/\beta>-1/2$\\
& & & & Saddle node:  $\lambda/\beta<-1/2$
\\
\hline
$F$ & $2\beta(\beta-\lambda)-9>0$ & $\frac{2\lambda+\beta}{\lambda-\beta}$ &  $-2/7<\frac{\lambda}{\beta}<1$ & Stable spiral:  region~VI  in Fig.~\ref{alpha3stability}\\
 & & &  & Saddle node or unstable node: otherwise \\
 \hline
\end{tabular}
\caption{Stability of critical points of Model I with $w=0$ and $\alpha=3$.}
\label{tab02}
\end{table}

\begin{figure}[!ht]
\includegraphics[scale=01]{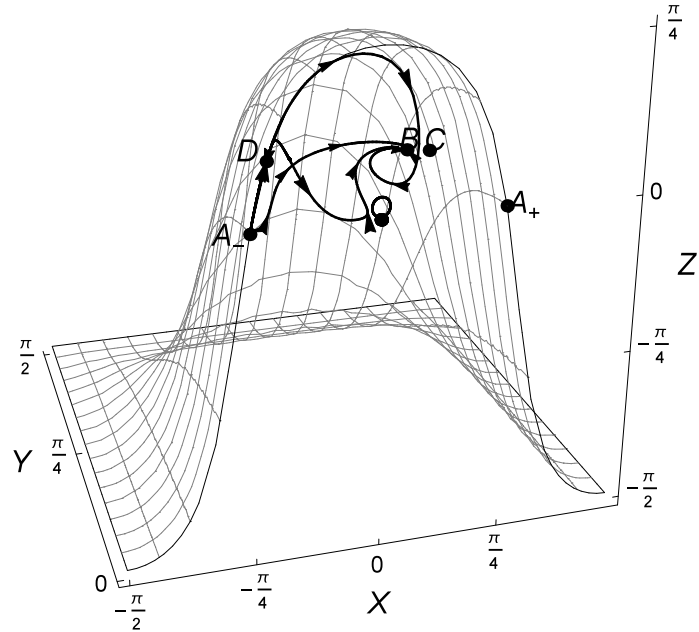}
\caption{Phase space showing trajectories of the dynamical system~(\ref{071})-(\ref{073}) when $\alpha=3,\beta=1,\lambda=2$ and $w=0$.}\label{alpha3beta1lambda2w0}
\end{figure}

\begin{figure}[!ht]
\includegraphics[scale=01]{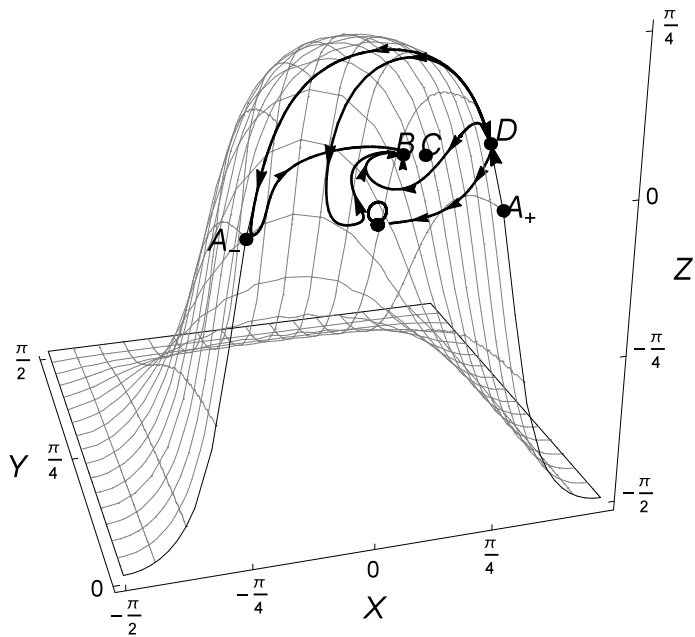}
\caption{Phase space showing trajectories of the dynamical system~(\ref{071})-(\ref{073}) when $\alpha=3,\beta=-1,\lambda=2$ and $w=0$.}\label{alpha3beta-1lambda2w0}
\end{figure}

The late time behaviour of this model is much more complex than the case $\alpha=1$. The qualitative late time behaviour of the phase space can be divided into essentially six regions of $\beta-\lambda$ parameter space. Two of these regions have two stable critical points, so the late time behaviour is dependent on the initial conditions. The remaining four regions have one stable point, corresponding to a global attractor. The different regions are plotted in Fig.~\ref{alpha3stability} displaying the corresponding critical points which are stable in that region. In order to plot trajectories in the phase space, we again compactify the phase space using the variables~(\ref{compactvars}).
We plot some example trajectories for two different choices of the parameters in Fig.~\ref{alpha3beta1lambda2w0} and Fig.~\ref{alpha3beta-1lambda2w0}.  The parameter choices for Fig.~\ref{alpha3beta1lambda2w0} correspond to the same parameter choices used in Fig.~\ref{alpha1beta1lambda2w0}, except now with $\alpha=3$ instead of $\alpha=1$, to enable us to see the effects of changing $\alpha$. In Fig.~\ref{alpha3beta1lambda2w0} the qualitative behaviour does not change much from the $\alpha=1$ case. Trajectories start at $A_{-}$, and end at the global attractor which is the point $B$. Many trajectories now pass near the point $D$, which for the choice of parameters used has an effective equation of state $w_{\rm eff}=4/3$, which is not physically viable. However in Fig.~\ref{alpha3beta-1lambda2w0}, where $\beta<0$, all the trajectories start at the point $D$, and end at the global attractor which is the scaling solution $B$.
This model never contains an early time accelerating solution, such a solution is required to describe an inflationary phase of the universe. The early time attractors are either $A_{\pm}$, $E$ or $D$, with $D$ being the only point that can accelerate for certain values of parameter space.
Unfortunately in the region in which $D$ describes an accelerating solution, the point is a saddle, so does not describe an early time inflationary universe.
Requiring special initial conditions, for which the universe remains in the vicinity of point $D$ for a sufficiently long time, one can however obtain a transient accelerating phase, which can still be used to describe early time inflationary stages.

The analysis of Model I with $\alpha=3$ shows the possible interesting applications at cosmological scales of this particular coupling between quintessence and dark matter.
One can in fact not only obtain late time accelerating attractors and scaling solutions, but also transient inflationary epochs useful for early time applications.
Moreover multiple late time attractors can be found for some values of the model parameters.
This is a feature which usually does not appear in quintessence models of dark energy and can interestingly be analysed with advanced dynamical systems methods such as bifurcation theory.
The complexity of Model I prevents a detailed study for every value of $\alpha$.
However the analyses we considered for the values $\alpha=1$ and $\alpha=3$ show that these interacting quintessence models can successfully be used to describe the dynamics of dark energy at late times.
Furthermore the presence of accelerating scaling solutions attracting the phase space trajectories at late times indicates a possible solution for the cosmic coincidence problem, as usually achieved introducing a coupling between dark energy and dark matter.

\subsubsection{Model II}

Now we will analyse the second model of the dynamical system~(\ref{071})-(\ref{073}) where we assume the interaction energy density to take the simple form $\rho_{\rm int}=\gamma \kappa \phi\rho$, where $\gamma$ is a dimensionless parameter. As before we can define an effective energy density and pressure, and this time the acceleration equation now reads
\begin{align}
\frac{\dot{H}}{H^2}=\frac{3}{2}[-(1+w)+(w-1)x^2+(1+w)y^2]\label{080}
\end{align}
and thus we can define the effective equation of state parameter in this model to be
\begin{align}
w_{\rm eff}=w-(w-1)x^2-(1+w)y^2
\end{align}
which we note is independent of the interaction energy density $z$. The critical points of this model are displayed in Tab.~\ref{tab:m2}.

\begin{table}[!ht]
\begin{tabular}{|c|c|c|c|}
  \hline
  Point & $x$ & $y$ & $z$ \\
  \hline \hline 
  $A_{\pm}$ & $\pm 1$ & 0 & 0 \\
  \hline
  $B$ & $\sqrt{\frac{3}{2}} \frac{(1+w)}{\lambda}$ & 
  $\sqrt{\frac{3}{2}} \frac{\sqrt{(1+w)(1-w)}}{\lambda}$ & $\frac{\lambda^2-3(1+w)}{\lambda^2}$ \\
  \hline
  $C$ & $\frac{\lambda}{\sqrt{6}}$ & $\sqrt{1-\frac{\lambda^2}{6}}$ &
  $0$ \\
  \hline
  $D$ & 0 & 0 & 1 \\
  \hline
\end{tabular}
\caption{Critical points of Model II.}
\label{tab:m2}
\end{table}

We see that all the critical points are independent of the parameter $\gamma$. We note that the origin of the phase space is not a critical point in this model. The existence and stability properties of the points are displayed in Tab.~\ref{tab04}. Depending on the value of $\lambda$, the system has either four or five critical points:
\begin{itemize}
\item {\it Point $A_{\pm}$}  These two points appeared in model I, and correspond to the scalar field kinetic energy dominated points with the effective equation of state reducing to that of a stiff fluid $w_{\rm eff}=1$. No acceleration is present at this point. These points are either unstable or saddle points depending on whether the absolute value of $\lambda$ is less than $\sqrt{6}$. 
\item {\it Point $B$} Unlike the canonical case when no interaction energy is present, this point exists for all values of $\lambda$. This solution is dominated by the energy density of the scalar field and the interaction energy density, and the universe behaves as if it is matter dominated at this point. The Jacobian matrix at this point has one zero eigenvalue, and two negative eigenvalues, and hence this point is not hyperbolic, meaning its stability properties cannot be determined by simply looking at the eigenvalues. To investigate its stability properties, one could use centre manifold theory, however here we will just examine this issue numerically. Based on numerical results, and by making a comparison with the corresponding point in the non-interacting scalar field model, we postulate that this point is stable if  $3(1+w)<\lambda^2$, and unstable otherwise.
\item {\it Point $C$} This point only exists when $\lambda^2<6$. It corresponds to a scalar field dominated universe, with both the matter and interaction energy densities vanishing. It is stable if $\lambda^2<3(1+w)$, and a saddle point otherwise. The effective equation of state parameter is $\lambda^2/3-1$ and hence the solution describes an accelerating universe if $\lambda^2<2$.  This point is the well known cosmological accelerated expansion driven by a sufficiently flat scalar field potential. 
\item {\it Point $D$} This point is entirely dominated by the interaction energy $z$. It exists for all $\lambda$, and always lies on the boundary of the phase space. It is a scaling solution with $w_{\rm eff}=w$, so the universe behaves as if it were completely matter dominated at this point.  This point is non-hyperbolic as one of the eigenvalues of its stability matrix is zero, however the matrix always has one positive eigenvalue so the point is unstable.
\end{itemize}

\begin{table}[!ht]
\begin{tabular}{|c|c|c|c|c|}
\hline
Point & Existence & $w_{\rm eff}$ & Eigenvalues & Stability \\
\hline\hline
$A_{-}$ & $\forall \lambda$ & 1 &$3-3w,3-3w,3+\sqrt{\frac{3}{2}}\lambda$ & Unstable node for $\lambda>-\sqrt{6}$ \\
& & & & Saddle node: otherwise \\
\hline
$A_{+}$ & $\forall \lambda$ & 1 & $3-3w,3-3w,3-\sqrt{\frac{3}{2}}\lambda$ & Unstable node: $ \lambda<\sqrt{6}$ \\
& & & & Saddle node: otherwise \\
\hline
$B$ & $\forall \lambda$ & $w$ & $0,-\frac{3}{4}(1-w)+\frac{3}{4\lambda}\sqrt{(1-w)(24(1+w)^2-\lambda^2(7+9w))}$ & Non-hyperbolic   \\
& & & $-\frac{3}{4}(1-w)-\frac{3}{4\lambda}\sqrt{(1-w)(24(1+w)^2-\lambda^2(7+9w))}$ & Unstable: $\lambda^2<3(1+w)$
\\
\hline
$C$ & $\lambda^2<6$ & $-1+\frac{\lambda^2}{3}$ & $\frac{\lambda^2-6}{2}, \lambda^2-3(1+w),$ & Stable node: $\lambda^2<3(1+w) $\\
& & & $\lambda^2-3(1+w)$ & Saddle node: $3(1+w)<\lambda^2<6$\\
\hline
$D$ & $\forall \lambda$ & $w$ & $0,\frac{3}{2}(w-1),\frac{3}{2}(w+1)$ & Unstable \\	
\hline
\end{tabular}
\caption{Stability of critical points of Model II.}
\label{tab04}
\end{table}

\begin{figure}[!htb]
\includegraphics[scale=01]{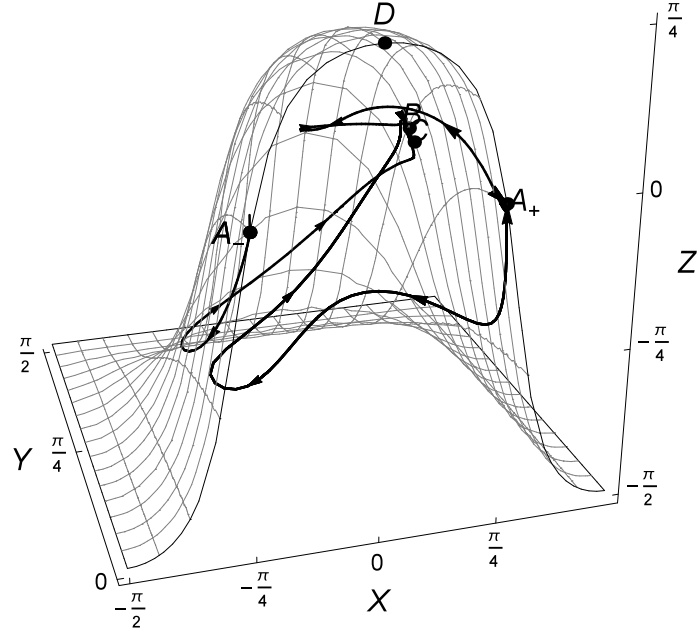}
\caption{Phase space showing trajectories of the dynamical system when $\gamma=1,\lambda=1.5$ and $w=0$.}\label{gamma1lambda1.5w0}
\end{figure}

\begin{figure}[!htb]
\includegraphics[scale=01]{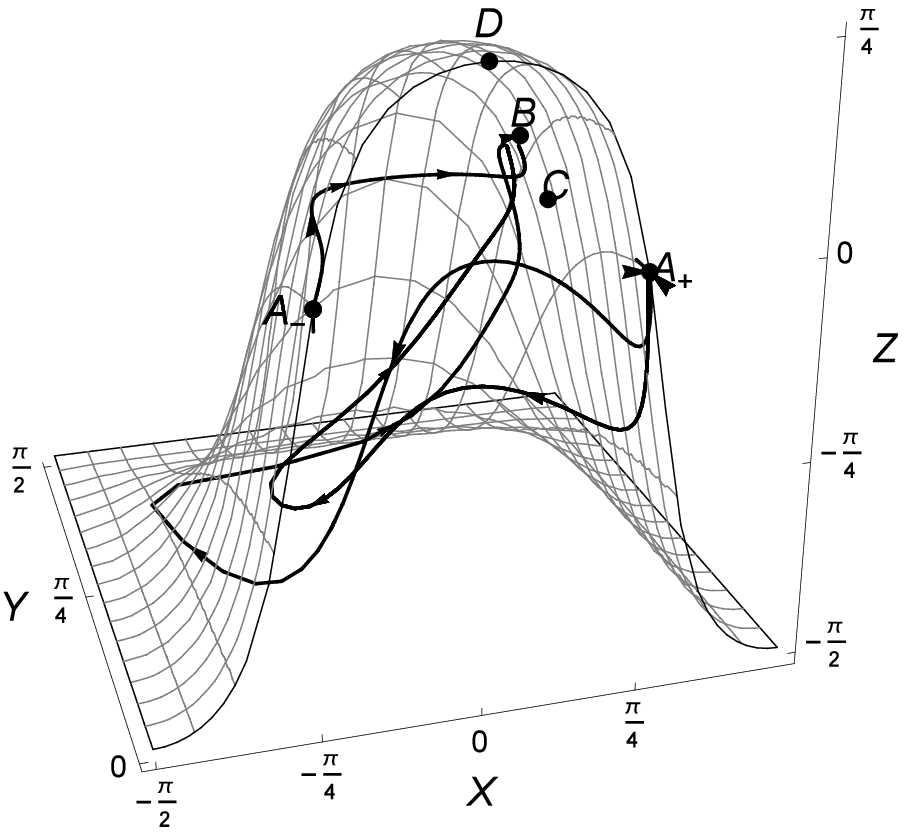}
\caption{Phase space showing trajectories of the dynamical system when $\gamma=1,\lambda=2$ and $w=0$.}\label{model2lambda2}
\end{figure}

\begin{figure}[!htb]
\includegraphics[scale=01]{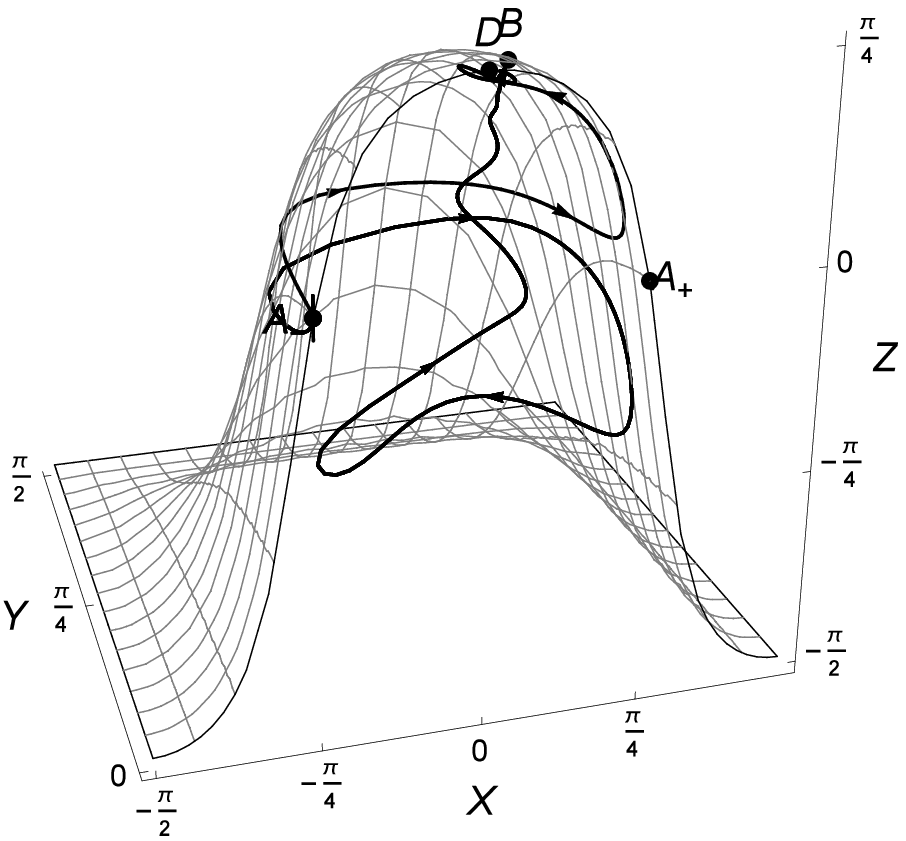}
\caption{Phase space showing trajectories of the dynamical system when $\gamma=1,\lambda=4$ and $w=0$.}\label{gamma1lambda4w0}
\end{figure}

In order to plot the trajectories, we again compactify the phase space using the variables~(\ref{compactvars}) as previously done for model I. We plot the phase space and some trajectories for three different values of $\lambda$, assuming a matter equation of state $w=0$ in Figs.~\ref{gamma1lambda1.5w0}, \ref{model2lambda2} and \ref{gamma1lambda4w0}.  

We see that this model qualitatively is almost identical to the case of the canonical scalar field when no interaction term is present. As seen in Figs.~\ref{gamma1lambda1.5w0}, \ref{model2lambda2} and \ref{gamma1lambda4w0}, all trajectories start at either of the stiff matter points $A_{\pm}$ and then evolve until they reach the global attractor which is either the point $B$ or $C$, the former corresponding to an effective dust (scaling) solution, while the latter corresponds to an accelerating universe solution if $\lambda^2<2$. The main difference between this model and the canonical scalar field is the origin is replaced by Point $D$, corresponding to an interaction dominated universe, and the point $B$ now exists for all values of $\lambda$. The matter dominated solution of the universe is thus represented by a phase in which the interaction term between dark energy and dark matter dominates. Though this phase can present the same dynamics of the $\Lambda$CDM model in the background evolution, differences might arise at the perturbation level. The analysis of cosmological perturbations as well as the comparison with observational data within the framework of these interacting dark energy models falls outside the scopes of the present work and will be left for future studies.

\section{Discussion and conclusion}
\label{sec:conclusion}

The main motivation of this work was to study interacting quintessence models which can be derived from a variational approach. Many interacting dark energy models considered in the past were constructed by adding suitable interaction terms at the level of the field equations. In many cases these terms could not be rigorously motivated from an underlining theoretical framework. Here we showed that one can use a rigorous Lagrangian formulation to define interacting quintessence models generally depending on one function $f(n,s,\phi)$ mixing the dynamical degrees of freedom of the matter fluid with the quintessence scalar field.

This Lagrangian formulation has been used to build first screening mechanisms for Solar System experiments and then interesting models of dark energy coupled to dark matter. As shown in Sec.~\ref{sec:small_scales}, new mechanisms capable of screening the effects of the scalar field at Solar System scales can be obtained. These solutions generalise the well-known chameleon mechanism which can now be derived from a variational approach. As an example of the new phenomenology that can be obtained in this manner, in Sec.~\ref{sec:small_scales} we considered a cosh-type of interaction proving that the scalar field can be highly massive at Solar System distances while being practically massless at cosmological scales.

The cosmological implications of our approach at the background cosmological level have then been studied in Sec.~\ref{sec:cosmology}. The main difference between the cosmology of our models and the one of previous models is that the matter conservation equation does not change in our case. Therefore, matter will always decay according to $\rho \propto a^{-3(1+w)}$ in agreement with the observations. However, the coupling does affect the Hubble constraint equation and therefore the total energy density depends on the coupling.

\begin{figure}[!hbt]
\includegraphics[width=0.48\textwidth]{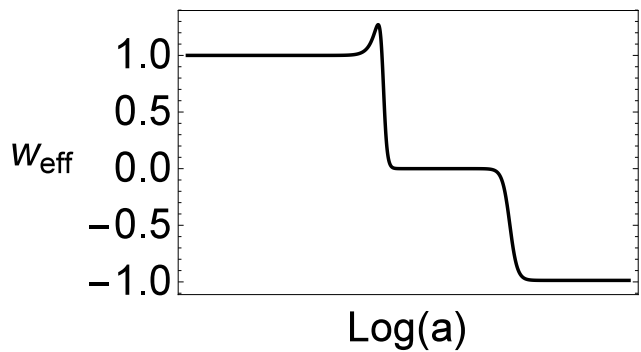}
\caption{The effective equation of state $w_{\rm eff}$ of model I with $\alpha=3$, $w=0$, $\beta=2$, and $\lambda=1/5$. We see a matter dominated epoch (corresponding to point $E$) for some time, followed by the dark energy dominated late time attractor (point $C$).}
\label{weffplot3}
\end{figure}

We proposed two simple interacting models, both of which have a cosmological three dimensional phase space, and can be studied efficiently using dynamical systems techniques. The natural phase space of these models is unbounded and we introduced new variables which make this phase space compact. The resulting models contain some very interesting phenomenology which we discussed in some detail. In particular for Model I we found late time accelerating solutions where the energy density of dark energy either dominates or scales according to the matter interacting energy density. This situation is similar to the one generally arising in standard interacting quintessence models and can be used to solve or at least alleviate the cosmic coincidence problem. As an example, Fig.~\ref{weffplot3} shows the behavior of the effective equation of state for one particular trajectory of model I. We note that at late times the universe evolves through a matter dominated epoch followed by an epoch of late time accelerated expansion, or simply dark energy. This is exactly the behavior observed for our universe, where a matter to dark energy transition must occur at late times. Note that for this particular trajectory the early time evolution does not correspond to the one we obtain from observations since, as shown by Fig.~\ref{weffplot3}, a stiff effective equation of state is always attained with even excursions to super-stiff region ($w_{\rm eff}>1$). This is a common feature of quintessence models with exponential potential \cite{Copeland:1997et} where the early time attractor is always represented by a scalar field dominated solution with a stiff equation of state, and not by a matter dominated solution or even an inflationary epoch as suggested by observations. In these situations the model is taken as an effective description valid at late times but not at early times.

\begin{figure}[!hbt]
\includegraphics[width=0.48\textwidth]{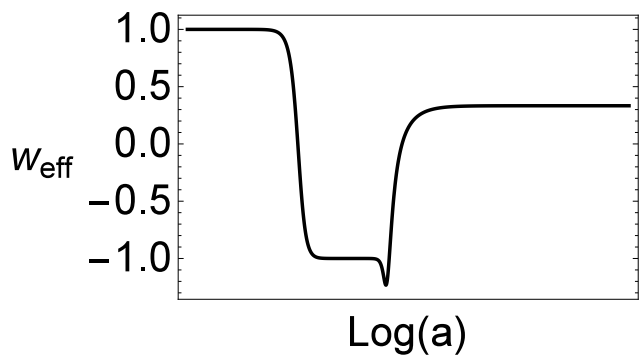}
\caption{The effective equation of state $w_{\rm eff}$ of model I with $\alpha=3$, $w=1/3$, $\beta=2\sqrt{3}$, and $\lambda=2$. After a stiff matter period, we see an inflationary era (corresponding to point $D$) for some time, before ending the late time attractor; the scaling solution point $B$.}
\label{weff9}
\end{figure}

Note however that the dynamics of model I (with $\alpha=3$) allows also for possible early time inflationary solutions described by Point~$D$.
This behaviour is neatly showed in Fig.~12 where the effective equation of state for a trajectory passing close to Point~$D$ has been plotted for $w=1/3$ and some particular values of the model parameters.
A transient early-time inflationary epoch, with $w_{\rm eff} = -1$ and a brief excursion into the phantom regime, is dynamically achieved after stiff domination, while Point~$B$ is the late time attractor describing a radiation dominated scaling solution, and not an accelerating solution.
Such a situation cannot be used to describe dark energy, but can interestingly be adopted to characterize an inflationary phase followed by a radiation dominated era, avoiding in this way the need for reheating, a remarkable result by itself.
Strong observational constraints are inevitably expected on such a model since the presence of the scalar field would be non negligible during radiation domination, Point~$B$ being a scaling solution.
Nevertheless this simple example shows how the scalar-fluid interaction presented in this work can be employed to produce new interesting phenomenology even at early times.

For a suitable choice of the model parameters, and for a fine tuning of initial conditions, such an inflationary solution might even be connected to a late time accelerating attractor through a transient matter dominated epoch. This is an interesting possibility since it would allow for a unified description of dark energy and inflation. An in depth analysis about the viability of this solution is outside the scope of the present paper and it will be left for future studies. Finally another interesting feature obtained within Model I is the possibility of multiple late time attractors which are theoretically and mathematically interesting to study.

Regarding Model II, at the background level it is almost (qualitatively) identical to the uncoupled standard scalar field cosmological models. However one expects differences at the perturbation level since in this model the scalar field has a non-negligible role during the matter dominated era, this being described by a scaling solution.

The framework we introduced to study interacting quintessence models is in fact even wider of the one considered here. In the present paper we focused on interactions of the form $f(n,s,\phi)$ where no spacetime derivatives appear. However, there is in principle no obstacle to consider functions which also depend on those derivatives. Under a theoretical perspective, such models are the next logical case to study and will be investigated in the second part of this work \cite{part2}. One could even consider radically different models where the entropy density $s$ becomes an increasing function of time instead of being a constant. Such models, taking into account the second law of thermodynamics, would contain a preferred direction of time and provide possible material for quantum gravity speculations.

In order to investigate the cosmological implications of all these models in more detail, one has to study their cosmological perturbations. This would clarify the issues of structure formation and stability, both of which are quite challenging to address in general interacting dark energy models. This is because at the level of perturbations spacetime derivatives of the coupling enter the perturbed field equations and in general there is no unique procedure to promote the background equations to fully covariant equations \cite{Koivisto:2005nr}. Our approach however makes this conceptually much easier since we have an underlying theory from which all equations are derived from first principles. In particular the fully covariant generalisations of the Einstein and Klein-Gordon field equations can be easily obtained, implying that the perturbations equations can be straightforwardly derived and analysed once an interaction has been specified at the level of the Lagrangian.

\acknowledgments

NT would like to thank Philippe Brax for useful discussions on the screening mechanisms of scalar fields.
The authors would like to thank the anonymous referee for the interesting feedback and useful comments on the paper.

\appendix

\section{Derivation of Eqs.~(\ref{051}) and (\ref{036})}
\label{appA}

In this appendix we will explicitly show the calculations around Eqs.~(\ref{051}) and (\ref{036}).
We start with the first one of Eqs.~(\ref{051}):
\begin{align}
U_\nu \nabla_\mu \tilde{T}^{\mu\nu} &= U_\nu \nabla_\mu \left[ \tilde{p} g^{\mu\nu} +(\tilde\rho + \tilde{p}) U^\mu U^\nu \right] \label{app:005} \\
	&= U^\mu \nabla_\mu\tilde{p} -U^\mu \nabla_\mu (\tilde\rho+\tilde{p}) -(\tilde\rho+\tilde{p})\nabla_\mu U^\mu + (\tilde\rho+\tilde{p}) U^\mu U_\nu \nabla_\mu U^\nu \label{app:001}\\
	&= -\frac{\partial\tilde\rho}{\partial n} \nabla_\mu (U^\mu n) -U^\mu \frac{\partial\tilde\rho}{\partial s} \nabla\mu s -U^\mu \frac{\partial\tilde\rho}{\partial\phi} \nabla_\mu\phi \label{app:002} \\
	&=-U^\mu \frac{\partial\tilde\rho}{\partial\phi} \nabla_\mu\phi \,, \label{app:006}
\end{align}
where in the line (\ref{app:001}) we used $U^\mu\nabla_\nu U_\mu=0$, obtained from the covariant differentiation of the constraint $U_\mu U^\mu=-1$, and in the line (\ref{app:002}) we used Eqs.~(\ref{019}), i.e.~$\nabla_\mu(n U^\mu)=0$ and $\nabla_\mu s =0$.

The second of Eqs.~(\ref{051}) is longer to verify.
Defining $\tilde\mu = \mu + \mu_{\rm int}$ and using again Eqs.~(\ref{019}) and $U^\mu\nabla_\nu U_\mu=0$, one finds
\begin{align}
h_{\mu\nu}\nabla_\lambda \tilde{T}^{\nu\lambda} &= (g_{\mu\nu}+U_\mu U_\nu) \nabla_\lambda\left( \tilde{p} g^{\nu\lambda} + (\tilde\rho+\tilde{p}) U^\nu U^\lambda \right) \label{app:007} \\
	&= h^\nu_\mu \nabla_\nu\tilde{p} + n \tilde\mu U^\lambda \nabla_\lambda U_\mu \\
	&= h^\nu_\mu \nabla_\nu\tilde{p} + n \left[ U^\lambda\nabla_\lambda(\tilde\mu U_\mu) - U_\mu U^\lambda \nabla_\lambda \tilde\mu \right] \\
	&= h^\nu_\mu \nabla_\nu\left( n\frac{\partial\tilde\rho}{\partial n}-\tilde\rho \right) + n \left[ 2 U^\lambda\nabla_{[\lambda}(\tilde\mu U_{\mu]}) + U^\lambda \nabla_\mu(\tilde\mu U_\lambda) - U_\mu U^\lambda \nabla_\lambda \frac{\partial\tilde\rho}{\partial n} \right] \\
	&= n \left[ 2 U^\lambda\nabla_{[\lambda}(\tilde\mu U_{\mu]}) + \frac{1}{n} h^\nu_\mu \nabla_\nu \left( \frac{\partial\tilde\rho}{\partial n} \right) -\frac{1}{n} h^\nu_\mu \nabla_\nu \tilde\rho - h^\nu_\mu \nabla_\nu\frac{\partial\tilde\rho}{\partial n} \right] \\
	&= n \left[ 2 U^\lambda\nabla_{[\lambda}(\tilde\mu U_{\mu]})+ h^\nu_\mu \frac{\partial\tilde\rho}{\partial n} \nabla_\nu n -\frac{1}{n} h^\nu_\mu \nabla_\nu\tilde\rho \right] \\
	&=  2 n U^\lambda\nabla_{[\lambda}(\tilde\mu U_{\mu]}) - h^\nu_\mu \frac{\partial\tilde\rho}{\partial\phi} \nabla_\nu\phi \,. \label{app:008}
\end{align}
In order to show that the first term in the last line above vanishes, we recall Eq.~(\ref{013}), namely
\begin{align}
	\tilde\mu\,U_\mu+ \varphi_{,\mu}+s\theta_{,\mu}+\beta_A\alpha^A_{,\mu} =0 \,.
	\label{app:009}
\end{align}
Using this we obtain
\begin{align}
	2 n U^\lambda\nabla_{[\lambda}(\tilde\mu U_{\mu]}) &= 2 U^\lambda \left[ \nabla_{[\lambda}\nabla_{\mu]}\varphi +s \nabla_{[\lambda}\nabla_{\mu]}\theta +\nabla_{[\lambda}\beta_A \nabla_{\mu]}\alpha^A + \beta_A \nabla_{[\lambda}\nabla_{\mu]}\alpha^A \right] \label{app:003} \\
	&= 2 U^\lambda \left( \nabla_{\lambda}\beta_A \nabla_{\mu}\alpha^A - \nabla_{\mu}\beta_A \nabla_{\lambda}\alpha^A \right) \label{app:004} \\
	&= 0 \,, \label{app:010}
\end{align}
where in the line (\ref{app:003}) we used the fact that covariant derivatives commute on any scalar, and in line (\ref{app:004}) we applied Eqs.~(\ref{004})-(\ref{007}).
We thus derive
\begin{align}
	h_{\mu\nu}\nabla_\lambda \tilde{T}^{\nu\lambda} = - h^\nu_\mu \frac{\partial\tilde\rho}{\partial\phi} \nabla_\nu\phi \,,
\end{align}
which corresponds to the second of Eqs.~(\ref{051}).

To prove Eqs.~(\ref{036}) one repeats the calculations (\ref{app:005})--(\ref{app:006}) with $T^{\mu\nu}$ instead of $\tilde{T}^{\mu\nu}$, obtaining simply
\begin{align}
	U_\nu \nabla_\mu {T}^{\mu\nu} = 0 \,,
\end{align}
because $\rho$ does not depend on $\phi$.
Analogously one repeats the calculations (\ref{app:007})--(\ref{app:008}) which yields
\begin{align}
	h_{\mu\nu}\nabla_\lambda {T}^{\nu\lambda} = 2 n U^\lambda\nabla_{[\lambda}(\mu U_{\mu]}) \,.
	\label{app:011}
\end{align}
Then, using again Eq.~(\ref{app:009}), and following the similar computations to (\ref{app:003})--(\ref{app:010}) one finds
\begin{align}
	h_{\mu\nu}\nabla_\lambda {T}^{\nu\lambda} = -2 n U^\lambda\nabla_{[\lambda}(\mu_{\rm int} U_{\mu]}) \,,
\end{align}
implying the validity of Eqs.~(\ref{036}).

\end{document}